# The list of possible double and multiple open clusters between galactic longitudes 240º and 270º

Juan Casado

Facultad de Ciencias, Universidad Autónoma de Barcelona, 08193, Bellaterra, Catalonia, Spain

Email: juan.casado@uab.cat

## Abstract

This work studies the candidate double and multiple open clusters (OCs) in the galactic sector from *l* = 240º to *l* = 270º, which contains the Vela-Puppis star formation region. To do that, we have searched the most recent and complete catalogues of OCs by hand to get an extensive list of 22 groups of OCs involving 80 candidate members. *Gaia* EDR3 has been used to review some of the candidate OCs and look for new OCs near the candidate groups. *Gaia* data also permitted filtering out most of the field sources that are not member stars of the OCs. The plotting of combined colour-magnitude diagrams of candidate pairs has allowed, in several cases, endorsing or discarding their link. The most likely systems are formed by OCs less than 0.1 Gyr old, with only one eccentric OC in this respect. No probable system of older OCs has been found.  Preliminary estimations of the fraction of known OCs that form part of groups (9.4 to 15%) support the hypothesis that the Galaxy and the Large Magellanic Cloud are similar in this respect. The results indicate that OCs are born in groups like stars are born in OCs.

## 1. Introduction

Open clusters are formed in giant molecular clouds and there is observational evidence suggesting that they can form in groups (Camargo et al. 2016). Galactic OCs are sometimes found in pairs, and the number of apparent pairs is significantly higher than would be expected if clusters were randomly distributed (e.g.Rozhavskii et al. 1976). Studies of clustering among OCs provide keys to understanding star formation in the Galactic disk and the subsequent dynamical evolution of OCs.

Nevertheless, until recently, h and $\chi$ Persei were the only confirmed physical double cluster known in our Galaxy (Vázquez et al. 2010). Conversely, roughly 10% of the known OCs in the Large Magellanic Cloud (LMC) seem to belong to pairs. Concerning the Galactic disk, the first estimations of the fraction of these binary clusters reached a maximum level of 20 % (Rozhavskii et al. 1976). Subramanian et al. (1995) estimated that about 8% of the OCs in the Galaxy appear to be members of binary systems. De La Fuente Marcos & de La Fuente Marcos (2009) argued that the real fraction is ca. 12%, similar to that in the LMC. However, the exact fraction remains an open question at present, which hopefully will be ascertained in the next future -at least in the solar neighbourhood- thanks to new precision *Gaia* data.

The second *Gaia* Data Release (*Gaia* DR2) provided precise full astrometric data (positions, parallax, and proper motions) and (1+2)-band photometry for about 1.3 billion stars (Gaia Collaboration 2018), starting a new era in precision studies of Galactic OCs (among other subjects). The recent third release of *Gaia* early data results (*Gaia* EDR3) improves, even more, the accuracy of the measurements for around 1.5 billion sources. The astrometric solution is accompanied by new quality indicators like the renormalised unit weight error (RUWE), which allow to easily filtering-out sources with imprecise data (Gaia Collaboration 2020).

Nowadays, virtually all new OCs are found through automatic data-mining techniques and, sometimes, machine-learning algorithms. Artificial intelligence techniques work out the plethora of data from large stellar databases and information provided by space missions, such as *Gaia* (e.g. Cantat-Gaudin et al. 2018; Castro-Ginard et al. 2018). However, these unsupervised searches by automatic algorithms do not detect a fraction of the existing OCs (Hunt & Refferet 2020). Thus, in a recent article (Casado 2020), the physical nature of a series of 20 new open clusters has been confirmed manually employing existing data from *Gaia* DR2. Remarkably, two of these unknown OCs, namely Casado 9 and 10, form a binary system. This unexpected finding motivated the present study on binary/multiple OCs in a 30º sector of the Galactic disc containing the new pair. The Galactic sector studied was also selected since it includes the Vela-Puppis star formation region, located roughly at $245^{o} < l < 265^{o}$ and $-15^{o} < b < -5^{o}$, which might represent the best example of the outcome of OCs formation in giant molecular clouds (Beccari et al. 2020). The limited sector surveyed allowed its manual study, going beyond the purely formal search of groups of OCs and examining each candidate group and its members individually, based on extensive available data.

The present study is organized as follows. In section 2 we define the methodology for the selection of candidate OCs belonging to groups detailed in Table 1 and provide a brief analysis from the literature of some groups in the studied area. Section 3 deals with the relationship between the age of OCs and their tendency to form part of binary or multiple systems. In section 4, some individual features of the proposed groups and member candidates are preliminary discussed, essaying to ascertain which OCs are more or less likely members. Section 5 extracts basic statistical information of the ensemble of results obtained. Finally, in Section 6 some conclusions are highlighted.

## 2. Selection methodology

As mentioned, *Gaia* source data allow the study of OC candidates and their eventual confirmation as physical systems. What is more, with the help of *Gaia* EDR3, it is possible to get rid of most of the contamination due to background stars that are not cluster's members, and the presence and structure of OCs are more clearly revealed in crowded fields (e.g. Fig. 4). The corresponding methodology and tools used here to find out and study new OCs have been described in Casado (2020). A similar method can be used to reveal binary or multiple systems in wider areas of the sky. A quality filter of RUWE < 1.4 has been routinely used to select the reliable *Gaia* EDR3 sources through this study (Gaia Collaboration 2020).

In addition to revising the existing literature on double and multiple OCs, our main raw materials to look for an extensive sample of group candidates have been the most recent and broadest catalogues of OCs (Kharchenko et al. 2013; Cantat-Gaudin et al. 2018; Bica et al. 2019; Liu & Pang 2019; Sim et al. 2019; Castro-Ginard et al. 2020; Cantat-Gaudin et al. 2020). For each of these catalogues, we have looked for close correlations between coordinates, PMs and parallax (or distance, *d*) through binary plots of each pair of parameters for all OCs within the studied area of the sky. For instance, when two OCs are found significantly closer than the average in the galactic longitude versus parallax plane, the rest of their astrometric measurements are cross-checked. If they seem to match, i.e. if there is some overlapping of the data considering their uncertainty interval at a 3σ level, both OCs are included in Table 1. The Table is later refined by retrieving the most comprehensive, accurate, and recent data on the individual OCs, from *Gaia* surveys when possible. When existing data are dubious, incomplete or inconsistent between different authors, the OCs are reexamined to manually obtain the corresponding parameters from *Gaia* EDR3. The resulting error intervals in Table 1 are not the standard deviations, but the absolute uncertainties encompassing all the sources likely to be members of each OC.

The primary requirement for any pair of OCs is that their proper motions (PMs) and parallaxes should match within the observational uncertainties. Following the criteria of previous studies, groups were further refined by ruling out OCs that are more than 100 pc away from any other members (Conrad et al. 2017; Liu & Pang 2019), assuming that they are at the mean distance of the group. Groups having differences in radial velocities (RVs) or tangential velocities significantly >10 km $s^{-1}$ (Conrad et al. 2017) were also discarded. However, a few exceptions are discussed on a case by case basis in section 4.

These order-of-magnitude thresholds are, of course, subjective. They are also conservative, i.e. they are not constraining enough to exclude any likely candidate pair of OCs. Further constraints to select the most likely groups are that the ratio Δplx/plx is less than 10% (the same criteria is used when comparing photometric distances) and that ΔPM/plx (or ΔPM *d*) is lower than 2 between each likely member of the grouping and at least another likely member, which implies that the differences in tangential velocities are also lower than 10 km $s^{-1}$.

To put in context, verify and further justify these selection criteria, we have compared the threshold values with the well-known h and $\chi$ Persei double cluster. From the angular distance among both OCs centres (27 arcmin) and assuming that they are at the same mean heliocentric distance (2.2 kpc; Cantat-Gaudin et al. 2020), an estimated distance among them of ca. 20 pc can be inferred, well within the 100 pc limit. The median reported RV for NGC 869 (h Persei) is -42.8 km $s^{-1}$ (Dias et al. 2002), and for NGC 884 ($\chi$ Persei) is -43 km $s^{-1}$ (Dias et al. 2002; Loktin & Popova 2017; Soubiran et al. 2018), which are practically coincident. The *Gaia* DR2 mean parallaxes are 0.399 mas and 0.398 mas, respectively (Cantat-Gaudin et al. 2020). Thus, Δplx/plx is less than 1%. Using proper motions from Cantat-Gaudin et al. (2020), both ΔPM/plx and ΔPM *d* are ≤ 0.2. Therefore, it is clear that the binary system formed by NGC 869 and NGC 884 fulfils all the adopted criteria for a well-behaved pair.

Liu and Pang (2019) searched for OC groups in their catalogue using the *Friend-of-Friend* algorithm. The criteria used was a linking length of 100 pc or less based on the star clusters' 3D positions only. We have revised their classification for the studied Galactic sector and some of their groups (or some of the OCs of certain groups) have not been included in Table 1 since their PMs are irreconcilable even considering the experimental uncertainty. Their group #50, containing 15 member candidates, resulted in two different groups in our classification (#11 and #19 in Table 1), taking into account the diverse PMs.

Conrad et al. (2017) have searched for cluster groups in the 6D parameter space, i.e. 3D position and 3D velocity. The criterion was again a linking length < 100 pc, and they used the clusters RVs to confirm the link when it is < 10 km $s^{-1}$. These authors have described three additional groups of OCs in our region of study (their groups #6 to #8), which are not included in Table 1 as their proximity (*d* < 0.4 kpc) does not allow us to see them as apparently close neither in their positions nor in their PMs. The groups contain 3, 2 and 4 OCs respectively.

As a rule, protoclusters, subclusters or embedded clusters (Bica et al. 2019), which are often being formed in small groups within the same giant molecular cloud (Camargo et al. 2016) or have an unusually small number of stars with high-quality *Gaia* data, are out of the scope of this study (see however comments on groups #6 and #15 in Section 4). That's the reason why another pair of clusters possibly related, namely [DBS2003] 19 (Liu & Pang 2019) and Bica 390 (Bica et al. 2019), are not included in Table 1. The same thing can be said about the dubious pair formed by Camargo 267 and the HD 77343 Group (Bica et al. 2019; Borissova et al. 2020). Finally, the candidate pair formed by Platais 9 and IC 2391 (Soubiran et al. 2018) has not been included in the present study because IC 2391 lies at *l* > 270º.

| Gr<br>1 | OC<br>2 | *l*<br>3 | *b*<br>4 | *plx*<br>5 | *d*<br>6 | $\mu_\alpha$<br>7 | $\mu_\delta$<br>8 | *R*<br>9 | *N*<br>10 | *Age*<br>11 | RV<br>12 | B<br>19 | C<br>20 | References and notes |
|---|---|---|---|---|---|---|---|---|---|---|---|---|---|---|
| # | Name | degree | degree | mas | kpc | mas yr$^{-1}$ | mas yr$^{-1}$ | arcmin | stars | Gyr | Kms$^{-1}$ | | | |
| 1 | NGC 2447 | 240.05<br>240.04 | 0.15<br>0.15 | <br>0.97 | 1.0<br>1.0 | -3.8<br>-3.6 | 3.9<br>5.1 | <br>12[a] | <br>731 | 0.58<br>0.58 | 22 | y | y | Conrad et al. 2017<br>Cantat-Gaudin+ 2020 |
| 1 | NGC 2448 | 240.76<br>240.85 | -0.26<br>-0.43 | <br>0.88 | 1.0<br>1.1 | -3.8<br>-3.4 | 4.7<br>2.9 | <br>16[a] | <br>121 | 0.02<br>0.10 | 24 | y | y | Conrad et al. 2017<br>Cantat-Gaudin+ 2020 |
| | | | | | | | | | | | | | | |
| 2 | UBC 461 | 241.44 | 2.58 | 0.29 | 3.2 | -2.7 | 3.1 | 11[a] | 72 | 0.05 | - | n | y | Cantat-Gaudin+ 2020 |
| 2 | UBC 462 | 241.45 | 2.19 | 0.29 | 3.4 | -2.7 | 3.0 | 6.6[a] | 8 | 0.04 | - | n | y | Cantat-Gaudin+ 2020 |
| 2 | Haffner 16 | 242.09 | 0.47 | 0.29 | 3.2<br>3.1 | -2.8 | 3.0 | 1,9[a] | 83 | 0.08<br>0.10 | <br>44 | y | y | Cantat-Gaudin+ 2020<br>Tarricq et al. 2021 |
| | | | | | | | | | | | | | | |
| 3 | Ruprecht 32 | 241.58 | -0.57 | 0.18 | | -2.0 | 3.1 | 2.6[a] | 13 | | - | y | y | Cantat-Gaudin+ 2020[f] |
| 3 | Casado 44 | 242.34 | 0.82 | 0.16 ± 0.05 | | -2.2 ± 0.2 | 3.2 ± 0.2 | 3.5 ± 0.5 | 19 | | - | | | This work[f] |
| | | | | | | | | | | | | | | |
| 4 | Casado 55 | 242.15 | -0.04 | 0.21 ± 0.06 | | $-2.5_5$ ±0.1 | 3.2 ± 0.2 | 2.0 ± 0.5 | 18 | | - | | | This work[f] |
| 4 | FSR 1315 | 243.20 | -0.63 | 0.21 | 4.5 | $-2.4_4$ | 3.1 | 4.8[a] | 38 | $0.08_1$ | 63[g] | y | y | Cantat-Gaudin+ 2020 |
| 4 | NGC 2453 | 243.27 | -0.94 | $0.20_5$ | 4.5 | -2.3 | 3.4 | 2.0[a] | 108 | $0.02_5$ | 34 –<br>86[g] | y | y | Cantat-Gaudin+ 2020 |
| 4 | Casado 34 | 243.55 | -0.40 | 0.20 ± 0.05 | | -2.5 ± 0.2 | 3.2 ± 0.2 | 2.0 ± 0.5 | 18 | | - | | | This work[f] |
| 4 | Casado 36 | 243.63 | 0.52 | 0.21 ± 0.06 | | -2.5 ± 0.2 | 3.2 ± 0.2 | 5 ± 1 | 33 | | - | | | This work |
| 4 | Casado 35 | 244.19 | -0.32 | 0.20 ± 0.06 | | -2.5 ± 0.3 | 3.1 ± 0.3 | 3.5 ± 0.5 | 30 | | - | | | This work |
| | | | | | | | | | | | | | | |
| 5 | UBC 637 | $242.29_1$ | $6.99_7$ | 0.18 | 4.7 | -2.1 | 2.0 | 1.9[a] | 28 | 1.7 | 100 | | | Cantat-Gaudin+ 2020[d] |
| 5 | UBC 638 | $242.30_3$ | $6.98_0$ | 0.09 | 6.1 | -2.1 | 2.0 | 0.7[a] | 10 | 1.6 | - | | | Cantat-Gaudin+2020[d,f] |
| | | | | | | | | | | | | | | |
| 6 | Waterloo 3 | 242.56 | 1.44 | 0.19 | 4.8 | -2.4 | 2.7 | | | 0.04-<br>0.32[g] | - | y | y | Cantat-Gaudin+ 2020<br>See UBC 465 |
| 6 | UBC 465 | 242.65 | 1.46 | 0.19 | 4.8 | -2.4 | 2.7 | 9.7[a] | 78 | 0.04 | - | n | y | Cantat-Gaudin+ 2020<br>Double core[b] |
| 6 | Haffner 19 | 243.07 | 0.53 | 0.17 | 5.5 | -2.5 | 2.5 | 1.0[a] | 39 | $0.03_6$ | 68[g] | y | y | Cantat-Gaudin+ 2020 |
| 6 | Haffner 18 | 243.16 | 0.45 | $0.18_5$ | 5.1 | -2.5 | 2.7 | 1.9[a] | 62 | $0.01_4$ | 60[g] | y | y | Cantat-Gaudin+ 2020[g] |
| 6 | UBC 639 | 243.22 | -0.36 | $0.19_5$± 0.05 | | -2.1 ± 0.1 | 2.8 ± 0.2 | 3.5±0.5 | 30 | - | 93 | | | This work[e] |
| | | | | | | | | | | | | | | |
| 7 | Casado 57 | 245.22 | 1.62 | 0.18 ± 0.05 | | -2.6 ± 0.2 | 3.2 ± 0.2 | 6 ± 1 | 42 | | - | | | This work |
| 7 | UBC 640 | 245.54 | 1.51 | $0.18_5$ | 4.7 | -2.5 | 3.2 | 6.1[a] | 39 | $0.08_7$ | - | n | y | Cantat-Gaudin+ 2020 |
| | | | | | | | | | | | | | | |
| 8 | Ruprecht 44 | 245.73 | 0.49 | 0.19 ± 0.04 | 4.6 –<br>5.8[g] | -2.4 ± 0.2 | 2.9 ± 0.2 | 5 ± 1 | 95 | 0.01[g] | 71 –<br>94[g] | y[p] | y | This work[e] |
| 8 | Ruprecht 43 | 245.93 | 0.36 | 0.18 ± 0.05 | 1.3[g] | -2.2 ± 0.2 | 2.9 ± 0.3 | 3.0 ± 0.5 | 30 | 0.25[g] | 104[g] | y | y | This work[e]<br>Double core |
| 8 | Casado 29 | 246.13 | 0.55 | 0.18 ± 0.05 | | -2.2 ± 0.2 | 2.5 ± 0.2 | 4.0 ± 0.5 | 29 | | - | | | This work |
| | | | | | | | | | | | | | | |
| 9 | NGC 2439 | 246.45 | -4.46 | 0.23 | 3.7 | -2.3 | 3.2 | 4.3[a] | 489 | 0.01 | 63 –<br>68[g] | y[p] | y | Cantat-Gaudin+ 2020 |
| 9 | Ruprecht 35 | 246.66 | -3.25 | 0.24 | 4.0<br>3.6 | -2.3 | 3.1 | 2.3[a] | 35 | 0.05<br>0.09 | 91[g] | y | y | Cantat-Gaudin+ 2020<br>Tarricq et al. 2021 |
| 9 | Haffner 15 | 247.95 | -4.16 | 0.24 | 3.8 | -2.2 | 3.3 | 1.9[a] | 151 | 0.03 | 51[g] | y | y | Cantat-Gaudin+ 2020 |
| 9 | Bochum 15 | 248.02 | -5.45 | 0.25 ± 0.03 | 2.8[g] | -2.3 ± 0.3 | 3.1 ± 0.3 | 3.5 ± 0.5 | 21 | $0.00_6$[g] | | y[p] | n | This work[e] |
| 9 | Casado 56 | 248.60 | -3.55 | 0.25 ± 0.02 | | -2.2 ± 0.2 | 3.1 ± 0.2 | 7 ± 1 | 17 | | - | | | This work[f] |
| 9 | UBC 238 | 248.95 | -4.08 | 0.24 | 3.5 | -2.1 | 3.1 | 10[a] | 120 | 0.02 | - | n | y | Cantat-Gaudin+ 2020 |
| 9 | MWSC 1331<br>(FSR 1351) | 249.29 | -4.80 | 0.26 | <br>1.6[g] | -2.1 | 3.1 | 20[c] | 217 | 0.25 -<br>0.16[g] | <br>142[g] | y[p] | n | Liu & Pang 2019 |
| 9 | FSR 1352<br>(MWSC 1334) | 249.30 | -4.71 | 0.24 | 3.6 –<br>1.6[g] | -2.0 | 3.1 | 2.0[a] | 16 | 0.01 -<br>0.19[g] | - | y | y | Cantat-Gaudin+ 2020 |
| | | | | | | | | | | | | | | |
| 10 | MWSC 1420 | 247.14 | 0.47 | 0.26 | 2.3[g] | -2.6 | 2.8 | 13[c] | 90 | 0.01 | - | y | n | Liu & Pang 2019 |
| 10 | FSR 1342<br>(MWSC 1391) | 248.14 | -1.22 | 0.23 | 4.4 | -2.4 | 2.7 | 2.2[a] | 66 | 0.03 | - | y | y | Cantat-Gaudin+ 2020 |
| 10 | Ruprecht 47 | 248.26<br>248.25 | -0.20<br>-0.20 | 0.23 | 3.7<br>3.6 | -2.5 | 2.7 | 2.5[a] | 52 | 0.03<br>0.03 | <br>74[g] | y | y | Cantat-Gaudin+ 2020<br>Tarricq et al. 2021 |
| 10 | LP 204 | 249.08<br>249.10 | 0.13<br>0.16 | 0.24<br>0.25 ± 0.07 | | -2.4<br>-2.5 ± 0.2 | 2.9<br>2.8 ± 0.2 | 18[c]<br>9 ± 1 | 161<br>87 | 0.56[g] | <br>61±4 | n | n | Liu & Pang 2019<br>This work |
| 10 | Ruprecht 48 | 249.13 | -0.63 | $0.23_5$ | 3.9 | -2.5 | 2.8 | 1.4[a] | 61 | 0.11 | - | y | y | Cantat-Gaudin+ 2020<br>Double core |
| 10 | Ruprecht 54 | 250.04<br>250.03 | 0.96<br>0.96 | 0.19<br>0.22 ±0.05 | 4.9 | -2.7<br>-2.7 ± 0.2 | 3.1<br>3.2 ± 0.2 | 2.2[a]<br>3.5 ± 0.5 | 66<br>67 | 0.28 -<br>0.82[g] | 69[g] | | | Cantat-Gaudin+ 2020<br>This work[e] |
| | | | | | | | | | | | | | | |
| 11 | UBC 7 | 248.87<br>248.74 | -13.35<br>-13.35 | 3.6 | <br>0.28 | -9.8 | 7.0 | 201[c] | 211 | 0,01 -<br>$0.03_2$ | <br>17 | n | y | Liu & Pang 2019<br>Tarricq et al. 2021 |
| 11 | Collinder 135 | 249.04<br>248.99 | -11.02<br>-11.20 | 3.3 | <br>0.29 | -10.1 | 6.2 | 87[c] | 92 | $0,05_5$<br>$0.02_7$ | <br>17 | y[p] | y | Liu & Pang 2019<br>Tarricq et al. 2021 |
| | | | | | | | | | | | | | | |
| 12 | Casado 38 | 249.07 | 0.13 | 0.20 ± 0.04 | | -2.8±0.1 | 3.6 ± 0.1 | 4.5 ± 0.5 | 23 | | - | | | This work |
| 12 | BH 19 | 249.88<br>249.88 | -0.02<br>-0.02 | 0.19<br>0.23 ± 0.05 | 4.6 | -2.9<br>-2.9 ± 0.2 | 3.5<br>3.6 ± 0.2 | 1.5[a]<br>2.5 ± 0.5 | 45<br>40 | 0.28 -<br>0.57[g] | 86[g] | y | y | Cantat-Gaudin+ 2020<br>This work[e] |
| 12 | UBC 471 | 250.38<br>250.38 | -0.55<br>-0.55 | 0.19<br>0.22 ± 0.03 | 4.4 | -2.8<br>-2.8 ± 0.2 | 3.8<br>3.7 ± 0.2 | 5.0[a]<br>8 ± 1 | 49<br>40 | 0.71 | <br>74±3 | n | y | Cantat-Gaudin+ 2020<br>This work[e] |
| | | | | | | | | | | | | | | |
| 13 | Casado 10 | 249.16 | -0.75 | 0.25 ± 0.05 | 3.6 | -3.1 ± 0.3 | 3.6 ± 0.2 | 3.5±0.5 | 27 | - | - | | | Casado 2020 |
| 13 | Casado 9 | 249.33 | -0.69 | 0.24 ± 0.05 | 3.7 | -3.2 ± 0.3 | 3.7 ± 0.2 | 5 ± 1 | 25 | - | - | | | Casado 2020 |
| | | | | | | | | | | | | | | |
| 14 | UBC 644 | 251.64 | -0.39 | 0.20 | 4.1 | -2.6 | 2.7 | 4.6[a] | 17 | 0.22 | - | n | y | Cantat-Gaudin+ 2020 |
| 14 | UBC 473 | 251.91 | -0.46 | 0.17 | 4.5<br>4.2 | -2.6 | 2.6 | 1.5[a] | 24 | 0.74<br>0.85 | <br>77 | n | y | Cantat-Gaudin+ 2020<br>Tarricq et al. 2021 |
| | | | | | | | | | | | | | | |
| 15 | MWSC 1460<br>[KPR2005] 45 | 253.74 | -0.22 | 0.28 ± 0.05 | 3.0[g] | -2.8 ± 0.2 | 3.1 ±0.3 | 5 ± 1 | 39 | $0.01_3$[g] | 43[g] | y | n | This work[e].<br>Double core |
| 15 | Majaess 95 | 254.06 | -0.09 | 0.35 ± 0.10 | - | -3.0 ± 0.4 | 3.3 ± 0.7 | 2.0 ± 0.5 | 19 | - | 31[g] | y | n | This work[e,f] |
| 15 | Casado 24 | 254.39 | -0.06 | 0.29 ± 0.05 | | -2.7 ± 0.2 | 3.1 ± 0.2 | 6 ± 1 | 35 | | - | | | This work |
| | | | | | | | | | | | | | | |
| 16 | Pismis 5 | 259.36 | 0.91 | 1.03 | 0.94 | -5.5 | 4.3 | 7.4[a] | 80 | 0.01 | - | y[p] | y | Cantat-Gaudin+ 2020 |
| 16 | Ruprecht 64 | 259.90 | 0.48 | 1.05 ± 0.07 | | -5.7 ± 0.4 | 3.7 ± 0.4 | 30 ± 3 | 159 | 0.07-<br>0.28[g] | 21-<br>27[g] | y | n | This work[e] |
| 16 | Collinder 197 | 261.53<br>261.52 | 0.94<br>0.96 | 1.02 | <br>0.93 | -5.8 | 3.9 | 19[c] | 167 | 0.01<br>$0.01_4$ | 36 –<br>21[g] | y[p] | y | Liu & Pang 2019<br>Tarricq et al. 2021 |
| 16 | MWSC 1579 | 262.55 | 1.50 | 1.02 | | -5.6 | 3.9 | 57[c] | 209 | 0.01 | 17[g] | y[p] | y | Liu & Pang 2019 |
| | | | | | | | | | | | | | | |
| 17 | Ruprecht 65 | 263.08 | -1.53 | 0.41 | 2.4 | -4.7 | 4.2 | 5.4[a] | 33 | $0.02_3$ | - | y | y | Cantat-Gaudin+ 2020 |
| 17 | UBC 245 | 263.82 | -2.58 | 0.42 | 2.3 | -4.6 | 4.3 | 7.1[a] | 52 | $0.03_4$ | - | n | y | Cantat-Gaudin+ 2020 |
| | | | | | | | | | | | | | | |
| 18 | Casado 28 | 263.22 | -2.37 | 0.51 ± 0.06 | | -5.5 ± 0.3 | 5.0 ± 0.3 | 10 ± 1 | 58 | | - | | | This work |
| 18 | NGC 2659 | 264.18 | -1.67 | 0.45 | 2.1 | -5.3 | 5.0 | 2.5[a] | 82 | 0.04 | 17[g] | y | y | Cantat-Gaudin+ 2020 |
| 18 | Casado 61 | 264.31 | -1.05 | 0.42 ± 0.08 | | -5.2 ± 0.3 | 5.2 ± 0.3 | 4.0 ± 0.5 | 25 | | - | | | This work |
| 18 | NGC 2645 | 264.79 | -2.90 | 0.52 | 1.8 | -5.9 | 5.1 | 2.6[a] | 36 | 0.03 | 17[g] | y[p] | y | Cantat-Gaudin+ 2020 |
| 18 | SAI 92<br>(=FSR 1436) | 264.91 | -2.87 | 0.54 ± 0.05 | <br>2.0 | -5.8 ± 0.4 | 5.1 ± 0.4 | 5 ± 1 | 91<br>148 | 0.10-<br>0.01[g] | -<br>- | y[p] | n | This work[e]<br>Buckner+ 2014 |
| 18 | LP 58 | 264.97 | -2.88 | 0.52 | | -5.8 | 5.1 | 46[c] | 482 | $0.02_3$ | - | n | n | Liu & Pang 2019 |
| 18 | Pismis 8 | 265.11 | -2,57 | 0.51 | | -5,7 | 4,9 | 15[c] | 64 | $0.04_7$ | 63[g] | y[p] | y | Liu & Pang 2019 |
| 18 | UBC 482 | 265.15 | -1.98 | 0.45 | 2.2 | -5.2 | 5.0 | 12[a] | 97 | 0.03 | - | n | y | Cantat-Gaudin+ 2020 |
| 18 | Gulliver 5 | 265.46<br>265.45 | -0.90<br>-0.92 | 0.41<br>0.44 ± 0.05 | 2.1 | -5.1<br>-5.0 ± 0.2 | 4.9<br>5.0 ± 0.2 | 6.4[a]<br>10 ± 1 | 15<br>86 | 0.04 | -<br>- | n | y | Cantat-Gaudin+ 2020<br>This work[e] |
| | | | | | | | | | | | | | | |
| 19 | UBC 480 | 264.86 | -10.76 | 2.1 | 0.48 | -5.4 | 5.7 | 5.6[a] | 13 | 0.2 | - | | | Cantat-Gaudin+ 2020[d] |
| 19 | LP 2388 | 265.12 | -10.69 | 2.1 | | -5.3 | 5.7 | 71[c] | 67 | 0.02 | 19[g] | | | Liu & Pang 2019[d] |
| | | | | | | | | | | | | | | |
| 20 | Bochum 7 | 265.11 | -1.94 | 0.17 ± 0.05 | 5.8 – 6.1[g] | -3.1 ± 0.2 | 3.6 ± 0.2 | 7 ± 1 | 63 | 0.01 – 0.03[g] | 49 | y[p] | n | This work[e]<br>Kharchenko et al.2013 |
| 20 | Casado 46 | 265.20 | -2.20 | 0.17 ± 0.05 | | -3.1 ± 0.2 | 3.6 ± 0.2 | 5 ± 1 | 58 | | - | | | This work.<br>Double core |
| 20 | Casado 49 | 266.06 | -2.08 | 0.18 ± 0.07 | | -3.4 ± 0.2 | 3.7 ± 0.2 | 3.5 ± 0.5 | 40 | | $51._7$ | | | This work |
| 20 | Teutsch 65 | 266.23 | -1.52 | 0.16 ± 0.06 | - | -3.1 ± 0.2 | 3.7 ± 0.2 | 3.0 ± 0.5 | 36 | - | - | y | n | This work[e] |
| 20 | Teutsch 222 | 266.49 | -2.00 | 0.17 ± 0.06 | - | -3.4 ± 0.2 | 3.7 ± 0.3 | 1.5 ± 0.5 | 22 | - | - | y[p] | n | This work[e] |
| 20 | Casado 48 | 266.99 | -2.57 | 0.17 ± 0.06 | | -3.3 ± 0.2 | 3.7 ± 0.2 | 2.5 ± 0.5 | 29 | | - | | | This work |
| | | | | | | | | | | | | | | |
| 21 | Casado 53 | 265.28 | -1.85 | 0.18 ± 0.07 | | -3.1 ± 0.2 | 3.8 ± 0.2 | 2.5 ± 0.5 | 23 | | - | | | This work |
| 21 | Casado 58 | 265.33 | -2.08 | 0.17 ± 0.06 | | -3.3 ± 0.2 | 3.9 ± 0.2 | 3.0 ± 0.5 | 27 | | - | | | This work.<br>Double core |
| 21 | Casado 59 | 265.96 | -1.65 | 0.16 ± 0.07 | | -3.3 ± 0.2 | 3.9 ± 0.2 | 3.0 ± 0.5 | 34 | | 70 | | | This work |
| | | | | | | | | | | | | | | |
| 22 | Casado 26 | 266.31 | -1.10 | 0.46 ± 0.07 | | -5.3 ± 0.2 | 4.8 ± 0.3 | 9 ± 1 | 57 | | - | | | This work |
| 22 | Ruprecht 71 | 266.38 | -1.93 | 0.49 | 1.8 | -5.2 | 4.6 | 3.5[a] | 39 | 0.02 | 39[g] | y | y | Cantat-Gaudin+ 2020 |
| 22 | UBC 483 | 267.41 | -1.53 | 0.52 | | -5.1 | 4.4 | 4.1[a] | 11 | | - | n | y | Cantat-Gaudin+ 2020[f] |
| 22 | Muzzio 1 | 267.96<br>268.00 | -1.32<br>-1.37 | 0.52 ± 0.04<br>0.52 | $1.8_6$ | -5.3 ± 0.2<br>-5.3 | 4.4 ± 0.2<br>4.2 | 8 ± 1<br>7.4[a] | 38<br>26 | $0.00_3$-<br>$0.00_8$[g] | -<br>136[g] | y[p] | y | This work[e]<br>Cantat-Gaudin+ 2020 |
| | | | | | | | | | | | | | | |
| 23 | Casado 51 | 266.95 | -2.16 | 0.18 ± 0.05 | | -3.6 ± 0.2 | 3.9 ± 0.2 | 3.0±0.5 | 28 | | $63._8$ | | | This work |
| 23 | Casado 50 | 267.55 | -2.20 | 0.17 ± 0.05 | | -3.6 ± 0.1 | $3.8_5$± 0.1 | 2.5±0.5 | 23 | | $53._8$ | | | This work |
| | | | | | | | | | | | | | | |
| 24 | ESO 312-4<br>(MWSC 1491) | 259.47 | -1.72 | 0.15 | 5.9 | -2.9 | 3.9 | $0.9_6$[a] | 30 | $0.04_4$-<br>0.56[g] | - | y | y | Cantat-Gaudin+ 2020 |
| 24 | Casado 60 | 259.72 | -1.62 | 0.20 ± 0.07 | | -2.9 ± 0.4 | 3.9 ± 0.3 | 2.0 ± 0.5 | 31 | | - | | | This work<br>Double core |

TABLE 1. Open cluster groups (Gr) and candidate member's properties within the Galactic sector 240º < *l* < 270º. Column headings: 1. Group number; 2. OC name; 3. Galactic longitude; 4. Galactic latitude; 5. Parallax; 6. Distance; 7. PM in right ascension; 8. PM in declination; 9 OC radius; 10. Number of member stars; 11. Age; 12. Radial velocity. Abbreviations: [a] radius containing 50% of members; [b] one of the cores coincides with Waterloo 3; [c] maximum cluster member's distance to average position; [d] possibly a unique OC (see text); [e] reexamined using *Gaia* EDR3 due to insufficient, imprecise or inconsistent reports; [f] not enough stars for a complete characterization; [g] see text; B19 listed by Bica et al. 2019; C20 listed by Cantat-Gaudin et al. 2020; y yes; n no; [p] protocluster or embedded cluster; + et al.

## 3. The age issue

According to Bhatia (1990), the lifetimes of binary clusters depend on their mutual separation, on the tidal force of the galaxy and, to a lesser extent, on encounters with giant molecular clouds. The theoretical models indicate binary cluster lifetimes ranging from a few Myr to 0.04 Gyr. Grasha et al. (2015) also found that in the NGC 628 galaxy, the clustering of stellar clusters decreases with cluster age for OCs older than 40 Myr. Table 1 shows that the grouped OCs are usually young, i.e. their age is < 0.1 Gyr. The most obvious interpretation is that they have been formed simultaneously from the same molecular cloud (de La Fuente Marcos & de La Fuente Marcos 2009) and that their age has not been enough to disperse them in the Galactic disk. Note that, in this case, the gravitational bound between the group members may be very loose. Some examples are discussed in the comments on each group in section 4 (e.g. Group 2 and Group 6).

The cited examples are consistent with the simultaneous formation hypothesis, which implies that the new OC member candidates of this group should most probably be young and have a similar chemical composition, and thus the colour-magnitude diagram (CMD) of them should be very similar and match when plotted in the same graph. To be precise, assuming that extinction and reddening are fairly constant within the limited area of a double cluster if both members have the same age and metallicity and are at the same distance, their CMD should fit as if they were part of the same OC (Casado 2020). Some examples are also given in Section 4.

Conrad et al. (2017) checked the age spread of potential candidates to distinguish between genuine groupings and chance alignments. However, the possible formation of the binary clusters via tidal capture or resonant trapping (de La Fuente Marcos & de La Fuente Marcos 2009) cannot be ruled out so far. Such pairs of OCs should share common kinematics but probably have rather different ages and chemical compositions and would tend to eventually merge (de Oliveira et al. 2000). Thus, a good match of their individual CMDs is not expected. A few of these cases are discussed when dealing with individual groups in Section 4 (e.g. Group 12).

Finally, when most of the OC members of multiple systems are young, but one or two of them are certainly older, there is a serious concern about their membership to the group, particularly for the most remote OCs ($plx \leq 0.2$) when the relatively large errors in parallax tend to confuse foreground and background objects in the same field. These exceptions will be also discussed on a case by case basis (Groups 9 and 10). Although this mixed formation scenario cannot be ruled out, its probability is expected to be so low that (most of) the older OCs have been discarded as member candidates. Accordingly, only one of the likely member candidates (MWSC 1331) has been reported to be significantly older than 0.1 Gyr by different authors, as we will see in the next Section.

## 4. Comments on individual groups

### *Group 1*

De La Fuente Marcos & de La Fuente Marcos (2009) and Conrad et al. (2017) proposed a binary OC formed by NGC 2447 and NGC 2448. Despite some similar parameters, *Gaia* data analyses suggest that both OCs may not be a physical pair taking into account their diverse proper motions. For instance, $\mu_\delta$ are 5.1 and 2.9 mas yr$^{-1}$ respectively (Cantat-Gaudin et al. 2020). The difference of parallaxes, higher than 10%, seems to be also significant. There is no consensus on the age of NGC 2447. For instance, Liu & Pang (2019) report 1.15 Gyr, in contrast to 0.58 Gyr reported by other authors (Conrad et al. 2017; Cantat-Gaudin et al. 2020). In any case, this OC seems much older than its alleged companion NGC 2448 (Table 1). Despite an age mismatch does not allow to rule out of a double system, all things considered, the physical nature of pair #1 appears to be dubious. The corresponding results in Table 1 exemplify the differences (and similarities) between previous data and *Gaia* DR2 for two well-known OCs.

### *Group 2*

All the reported parameters of this trio of OCs seem to nicely match. The three OC member candidates have been previously studied and are relatively young. The only shadow is that the distance between Haffner 16 and UBC 462 appears to be very near the adopted limit of 0.1 kpc. Anyhow, this is considered a probable physical system.

### *Group 3*

The pair formed by Ruprecht 32 and the previously unknown cluster Casado 44 is questionable. Although most of the tabulated data are compatible considering the experimental uncertainty, the derived distance between them (0.16 kpc) exceeds the criteria of proximity usually adopted in the literature. Their small number of member stars (Table 1) precludes a complete characterization so far (including important parameters such as the age), but their most salient features (core star densities, preliminary CMDs and the presence of nebulosity only in Ruprecht 32) suggest that their association may be only apparent. For instance, the star densities within the radius containing 50% of the members are ~1 and ~5 arcmin$^{-2}$, respectively. The absence of RV measurements avoids the checking out of this parameter. Therefore, this is considered a dubious candidate group.

### *Group 4*

This candidate is a probable group, taking into account the overall consistency of the individual OC measurements in Table 1. However, the membership of NGC 2453 is dubious. It appears much brighter, star-rich and denser, and its mean PMs do not coincide with all the rest of the group members. Although both NGC 2453 and FSR 1315 seem to be young, the ages reported by the same authors are somewhat different (Table 1), and, accordingly, the fitting of their CMDs is not perfect. Unfortunately, the RVs do not allow reaching a definite conclusion. The reported RVs of NGC 2453 range from 34 km s$^{-1}$ (Dias et al. 2002) to 86 km s$^{-1}$ (Soubiran et al. 2018), and the reported RV for FSR 1315, 63 km s$^{-1}$ (Soubiran et al. 2018; Tarricq et al. 2021), lies within such interval.

Casado 55 is very near to the limit of distance to FSR 1315, its closest OC in the group, and its membership is uncertain. On the other hand, all the obtained data on Casado 34 and 36 suggest that they are probable group members. No RV measurements have been found for the stars of the new OC member candidates.

### *Group 5*

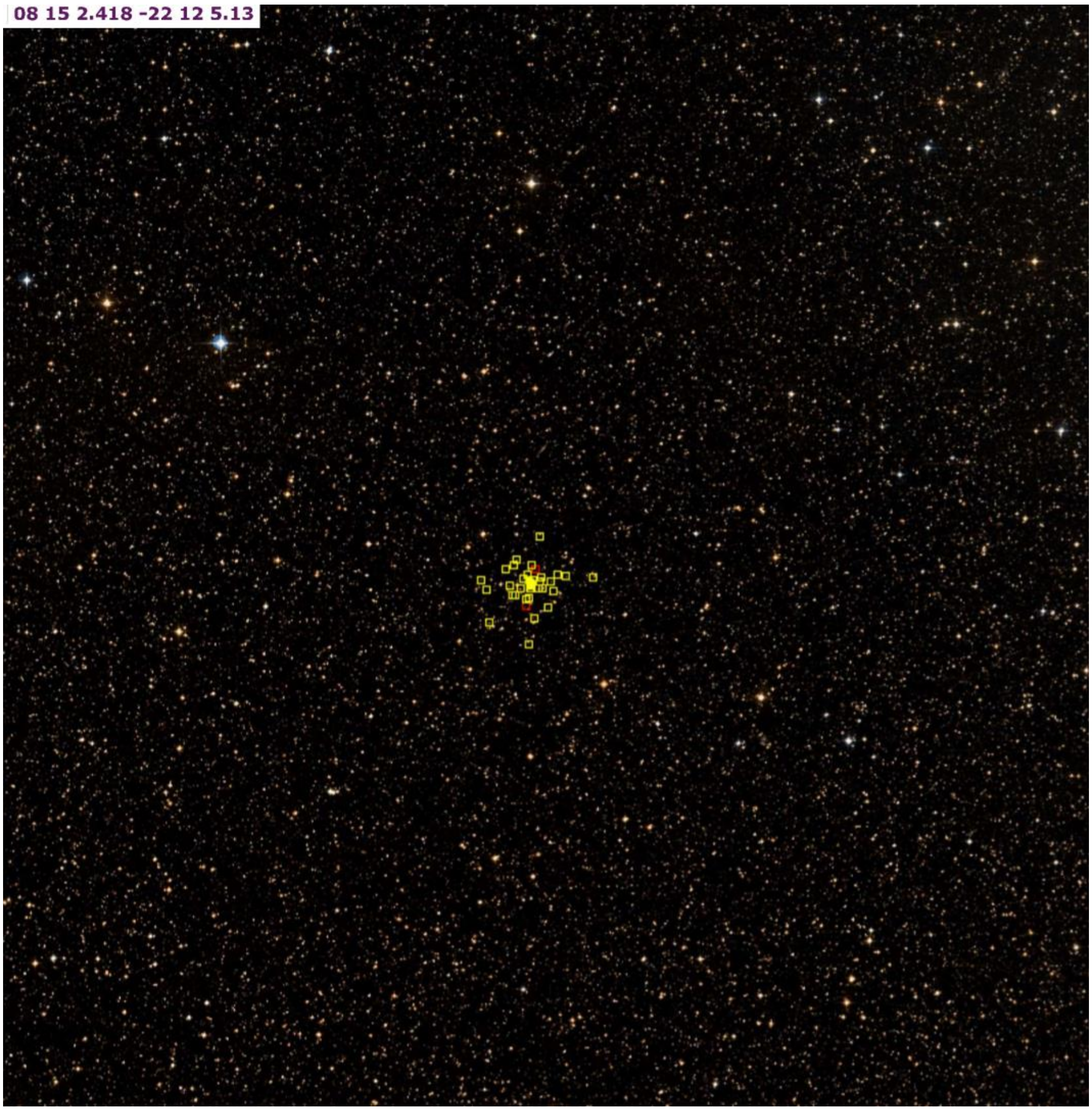

Figure 1. Probable member stars of the system UBC 637-638 (yellow squares). The reported centres of each OC are marked by red squares. RA and Dec of the image centre are given in the inset.

The case of UBC 637 and 638 is unusual. All the data, i.e. positions, PMs, ages and sizes, are concordant except the mean parallaxes and distances. Moreover, the parallax of UBC 638 is at odds with its reported distance. The angular distance between their centres (1.2 arcmins) is smaller than the reported radius of at least one of them (UBC 637). Furthermore, the spatial distribution of the member stars appears to share a common core that corresponds to an apparent overdensity of stars between the reported centroids of both objects (Fig. 1). If one accepts that the cited differences are within the observational uncertainty, which is consistent with most of the individual errors for so small parallaxes, the most plausible conclusion would be that both objects are the same OC. In such a case, the mean parallax of the 34 most probable members of the joint OC indicates a distance of 6 kpcs, in good agreement with one of the reported distances.

On the other hand, if significantly different distances were confirmed, then there is no binary system of OCs. But in such a case, the perfect fitting of the CMD for the ensemble of members of the system, shown in Figure 2, would be very hard to justify.

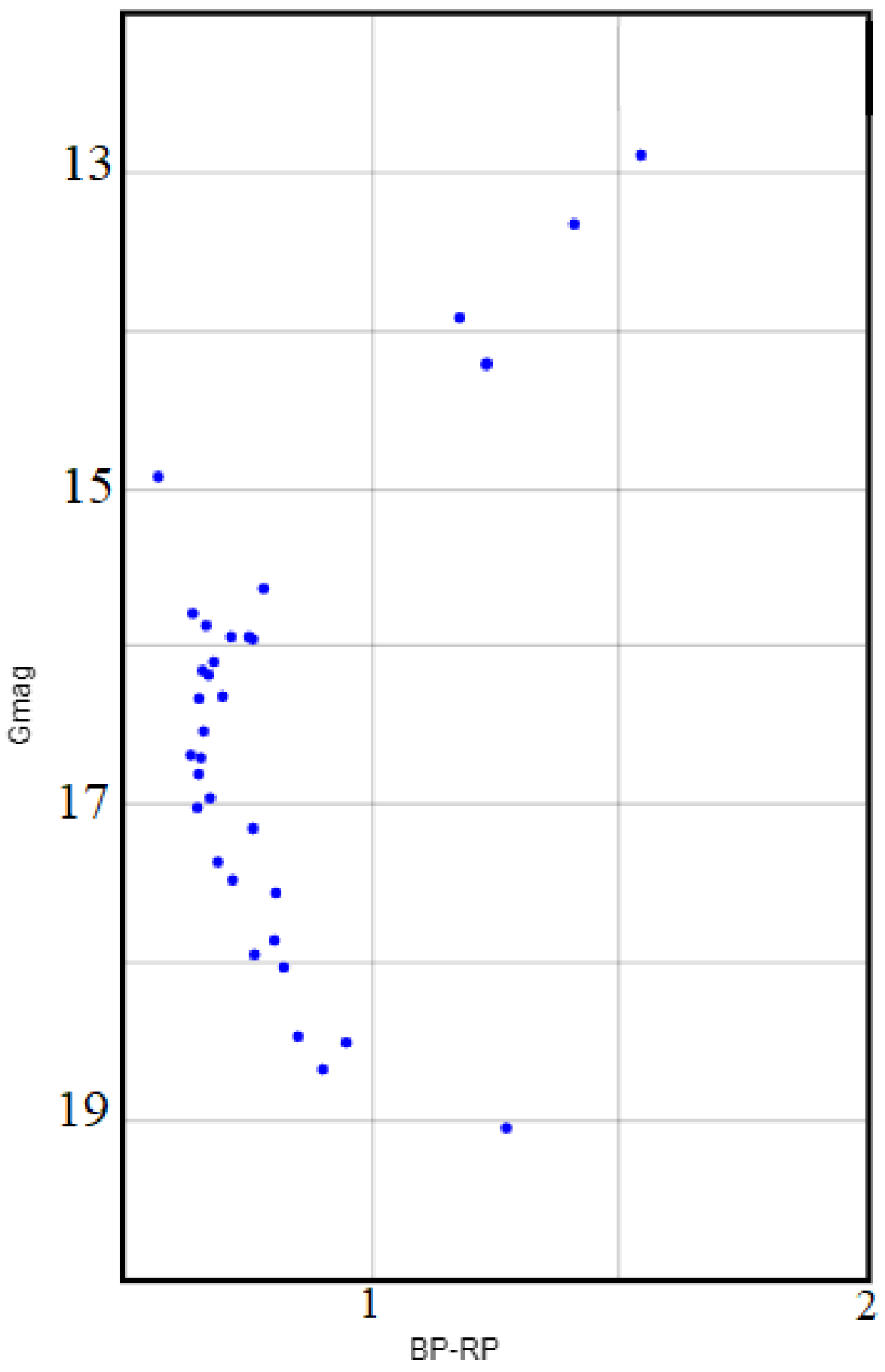

Figure 2. Combined CMD plot of the system UBC 637 + 638. Constraints: *plx* 0.08 to 0.20 mas; $\mu_\alpha$ -2.0 to -2.4 mas yr$^{-1}$; $\mu_\delta$ 1.7 to 2.3 mas yr$^{-1}$; *r* = 3 arcmin.

### Group 6

Two subgroups of close OCs can be discerned. One is formed by Waterloo 3 and UBC 465. The second is the system of Haffner 18 and 19, still in their parent cloud. Both subgroups are ca. 0.09 kpc apart, assuming that they are at the same heliocentric distance.

Waterloo 3 seems to be a part of UBC 465, which is a double core OC. However, its membership to this multiple group should be confirmed since a markedly older age (0.32 Gyr) has been also reported (Kharchenko et al. 2013). Such a wide span in reported ages for an OC is not an exceptional situation: at least 30% of OC ages have errors > 50%, especially for young OCs due to the low visibility of the turn-off point (Paunzen & Netopil 2006). In particular, the Kharchenko et al. (2013) catalogue is not very suitable as a source of ages for young OCs, because the parameters reported in that catalogue (except for proper motions) are based on near-IR photometry (2MASS), and the corresponding CMDs have low age sensitivity in this age interval. As a result, this catalogue is the source of many large discrepant ages in the identified groups. Anyway, the presence of the associated reflection nebula ESO 494-1 suggests that Waterloo 3 may well be a young cluster and may belong to the group. The rest of the reported ages of all members correspond to young OCs.

Vázquez et al. (2010) concluded that Haffner 18 and 19 have incompatible distances and ages to share a common origin. These authors reported distances to Haffner 18 and 19 of 4.5 ± 0.4 and 6.4 ± 0.65 kpc, respectively. Conversely, *Gaia* derived parallaxes and distances (Table 1) are compatible. Other reported distances to Haffner 18 in the literature range from 4.6 kpc (Glushkova et al.1997) to 6.0 kpc (Loktin & Matkin 1994), while reported distances to Haffner 19 range from 4.6 kpc (Loktin & Popova 2017) to 6.4 kpc (Dias et al. 2002). The distances in Table 1 seem to be a good compromise for the ensemble of these data. Concerning the ages of Haffner 18 and 19, both are young clusters according to most reports. Vázquez et al. (2010) reported 0.18 ± 0.02 Gyr for Haffner 18 and 0.005 Gyr for Haffner 19. However, the rest of the ages for Haffner 18 range from 0.001 Gyr (Ahumada & Lapasset 1995) to 0.016 Gyr (Dambis 1999), and most reports agree in 0.008 Gyr (Dias et al. 2002; De Marchi et al. 2006; van den Bergh 2006; Ahumada & Lapasset 2007). All of them are incompatible with the age in Vázquez et al. (2010). Ages in the literature for Haffner 19 range from 0.004 Gyr (e.g. Dias et al. 2002) to 0.036 Gyr (Cantat-Gaudin et al. 2020). This interval encompasses the age in Vázquez et al. (2010). Moreover, RVs are similar: 60 km $s^{-1}$for Haffner 18 (Kharchenko et al. 2013; Loktin & Popova 2017) and 68 km $s^{-1}$for Haffner 19 (Dias et al. 2002; Kharchenko et al. 2013; Loktin & Popova 2017). All in all, the existing data rather suggest a common origin between both close OCs. The embedded clusters Haffner 18b and 18c (Bica et al. 2019) appear to be also associated with the group. The NGC 2467-east IR cluster could also be a member of this complex system, considering its close position and a concordant RV of 60 km $s^{-1}$ (Kharchenko et al. 2013). An OC remnant candidate having seven stars in a radius of 1 arcmin at galactic coordinates 242.98, 1.62 could also belong to the group. If right, the group population would increase to at least seven members.

On the other hand, The UBC 639 membership is doubtful. It is quite apart from the other group members and its parallax and PMs are only marginally compatible with the rest. Moreover, the RV of one of its likely stars, 93 km $s^{-1}$, advises to rule out this OC of the group.

### *Group 7*

These two OCs are rather close to each other. Their *Gaia* results are equivalent but there are no available RVs for the member stars. Our approach has been to obtain a CMD of the smaller field comprising both OCs and some halo stars. If both clusters have the same age and metallicity and are at the same distance, both of their CMD should fit (Casado, 2020). The combined CMD (Figure 3) shows a broad main sequence with a few outliers, indicating that both are young OCs of compatible age (Table 1) and origin. The apparent close superposition of two sequences is due to the slightly different characteristics of each OC. Therefore, the real existence of this pair is plausible.

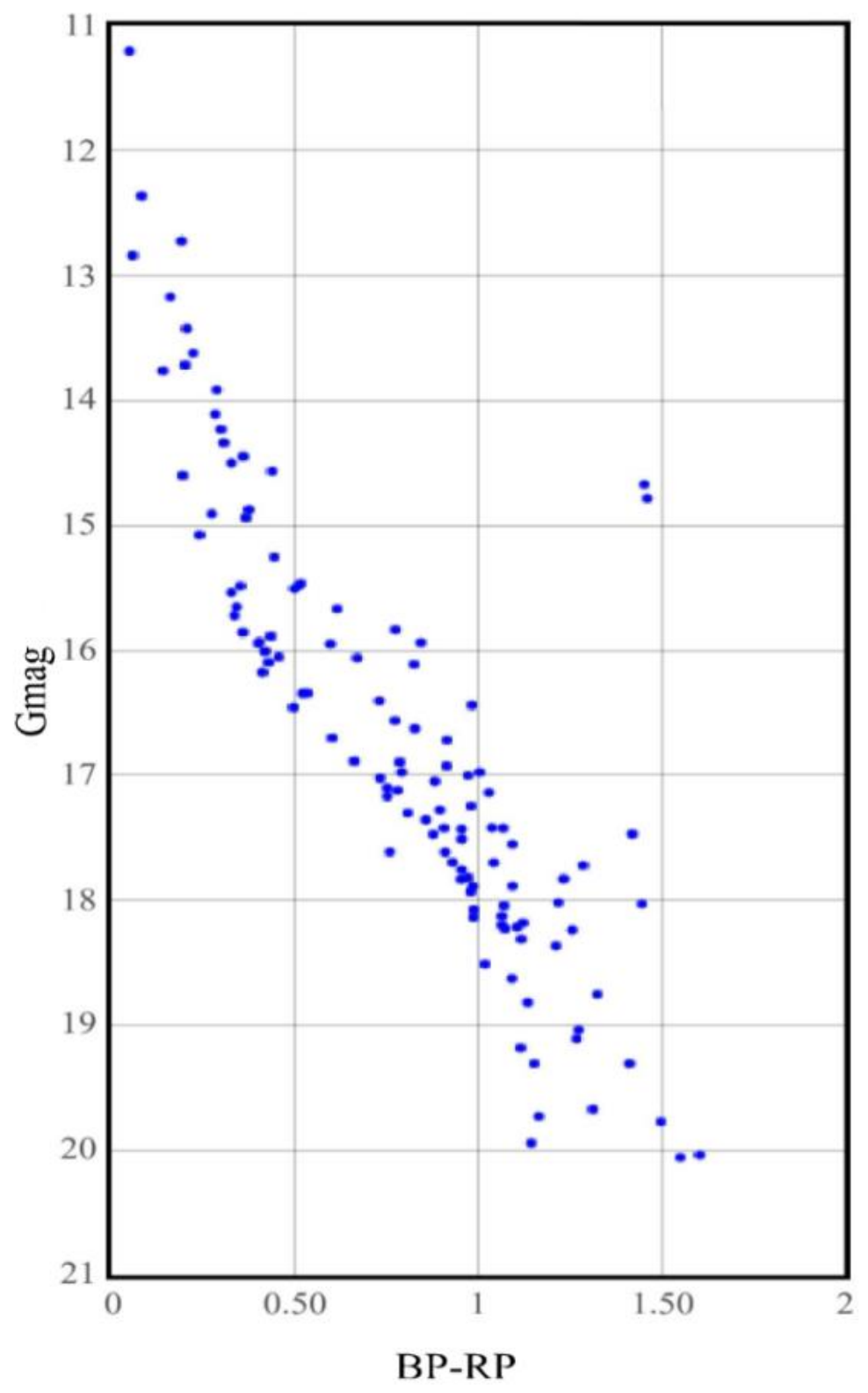

Figure 3. Combined CMD plot of the binary system formed by UBC 640 and Casado 57. Constraints: *plx* 0.15 to 0.22 mas; $\mu_\alpha$ -2.3 to -2.8 mas yr$^{-1}$; $\mu_\delta$ 3.0 to 3.4 mas yr$^{-1}$; *r* = 14 arcmin.

## *Group 8*

The distances of Ruprecht 44 in the literature range from 4.6 kpc (Loktin & Popova 2017) to 5.8 kpc (Cantat-Gaudin et al. 2020), and there is a broad consensus on its age: ~ 0.01 Gyr (Loktin & Matkin 1994; Dias et al. 2002; Ahumada & Lapasset 2007; Cantat-Gaudin et al. 2020). So far, the only reported RV for Ruprecht 44 was 94 ± 30 km s$^{-1}$ from a single star (Dias et al. 2002). We have derived a mean RV of 71 ± 10 km s$^{-1}$ from published data on 12 B-type stars of the OC (Garmany et al. 2015).

Considering only the literature data, it would seem dubious if Ruprecht 43 is related or not to Ruprecht 44. Both clusters positions are close, and their astrometric *Gaia* data nicely agree, but the reported age of Ruprecht 43 is at least 0.25 Gyr old (Kharchenko et al. 2013), and its RV would be at least 104 ± 2 km s$^{-1}$ (Soubiran et al. 2018). Nevertheless, we have to be cautious with pre-*Gaia* estimations of some parameters (Paunzen & Netopil 2006). For instance, the previous maximum distance, 1.3 kpc (Dias et al. 2002), is strikingly at odds with the mean parallax obtained from *Gaia* EDR3. In any case, the CMD of both OCs fit reasonably. So, Ruprecht 43 may well be a member of this group. This OC has a secondary core at galactic coordinates 246.12, 0.36, 11 arcmin away from the first one (Hunt & Reffert 2020), which could also be an associated member.

The previously unknown OC Casado 29 might form a pair with Ruprecht 44 considering its astrometric measurements from *Gaia* EDR3, excluding the marginal match of $\mu_\delta$. On the other hand, both CMDs closely fit.

The similar parameters of groups 6 and 8 suggest that both groups could result from the breakup of a primordial one.

## *Group 9*

MWSC 1331 (or FSR 1351) and FSR 1352 form probably a pair. As shown in Figure 4, their positions are very close, and their Gaia astrometric data fit well (Table 1). Although the reported ages for FSR 1352 vary widely from ca. 0.01 Gyr (Cantat-Gaudin et al. 2020) to 0.19 Gyr (Kharchenko et al. 2013), there is internal consistency in the ages of both OCs obtained by the same authors. Ages reported for MWSC 1331 are 0.16 Gyr (Kharchenko et al. 2013) and 0.25 Gyr (Liu & Pang 2019). Something similar happens with the reported distances. They span from 3.6 kpc (Cantat-Gaudin et al. 2020) to 1.51 kpc (Kharchenko et al. 2013) for FSR 1352, but the MWSC 1331 distance reported by Kharchenko et al. (2013) is very similar: 1.58 kpc. The shorter distances probably suffer from some systematic error considering the *Gaia* parallaxes (Table 1). Flawed distances could also result in flawed ages. On the other hand, the only RV of MWSC 1331, 142 km $s^{-1}$ (Kharchenko et al. 2013), is at odds with the RVs of the rest of the group members, but this quantity comes from merely one star.

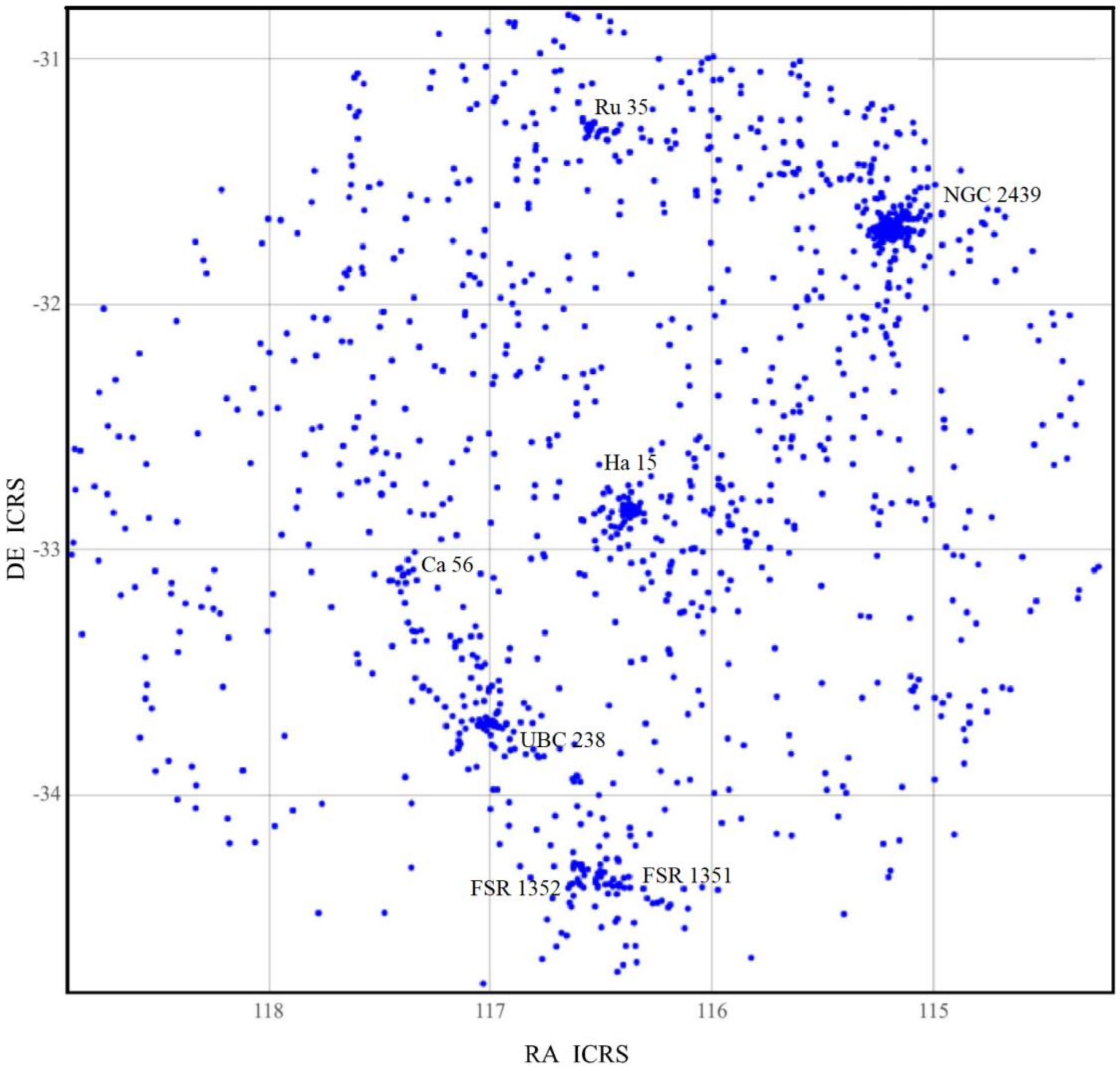

Figure 4. Chart of selected *Gaia* EDR3 sources near the group containing NGC 2439 (group 9). Most noise from non-member field stars has been removed by applying the constraints: *plx* 0.23 to 0.27 mas; $\mu_\alpha$ -1.9 to -2.3 mas $yr^{-1}$; $\mu_\delta$ 3.0 to 3.4 mas $yr^{-1}$.

The whole group appears to be dominated by NGC 2439, the most populated OC (Fig. 4). All its known parameters (Table 1) agree with the rest of the candidate members. The reported RVs are between 63 km $s^{-1}$ (Kharchenko et al. 2013) and 68 km $s^{-1}$ (Conrad et al. 2017; Soubiran et al. 2018). Ruprecht 35 data are also well-matched with the rest of the OCs in this group, excluding its RV, 91 km $s^{-1}$ (Soubiran et al. 2018), which comes from only one star. The only reported RV of Haffner 15 is 51 ± 8 km $s^{-1}$ (Garmany et al. 2015) and seems compatible with its membership to the group of NGC 2439. A small clump (*r* ~ 1 arcmin) of nine stars, including the blue supergiant HD 63804, appears to be also part of the group. Bridges of halo stars seem to connect some of the OCs in the group, such as UBC 238 and Casado 56 via the just mentioned clump (Fig. 4). Similar tidal bridges or tails have been found in some model simulations of interacting binary OCs (e.g. De La Fuente Marcos & de La Fuente Marcos 2010). These features and the angular distances among the OCs, allow discerning two subgroups: a first one formed by NGC 2439, Haffner 15 and Ruprecht 35, and a second one comprising the rest of the candidate members.

On the other hand, the membership of Bochum 15 is dubious. Although the astrometric data fit well, the photometric distance, 2.8 kpcs (Dias et al. 2002; van den Bergh 2006; Loktin & Popova 2017), seems too small. There is no RV measurement to check out its membership.

### Group 10

We have revisited LP 204 since it is considered an OC candidate that requires further confirmation (Liu & Pang 2019). Our results are compatible with those previously reported (and with the rest of the likely group members), excluding the radius and the number of potential star members (Table 1). One of these stars has RV = 61 ± 4 km $s^{-1}$. The CMD confirms the real existence of this OC, which includes a compact core at its centre. The reported age for LP 204 is 0.56 Gyr. Thus, the membership of this older cluster to a group of mostly young OCs is questionable. Nevertheless, the presence of the interstellar matter cloud BRAN 101 within the apparent limits of LP 204 suggests that this OC could be younger and its age requires confirmation.

The membership of MWSC 1420 is doubtful, Both the parallax and the reported distance, 2.3 kpc (Dias et al. 2002), suggest that it is closer to the sun than the rest of the candidate members.

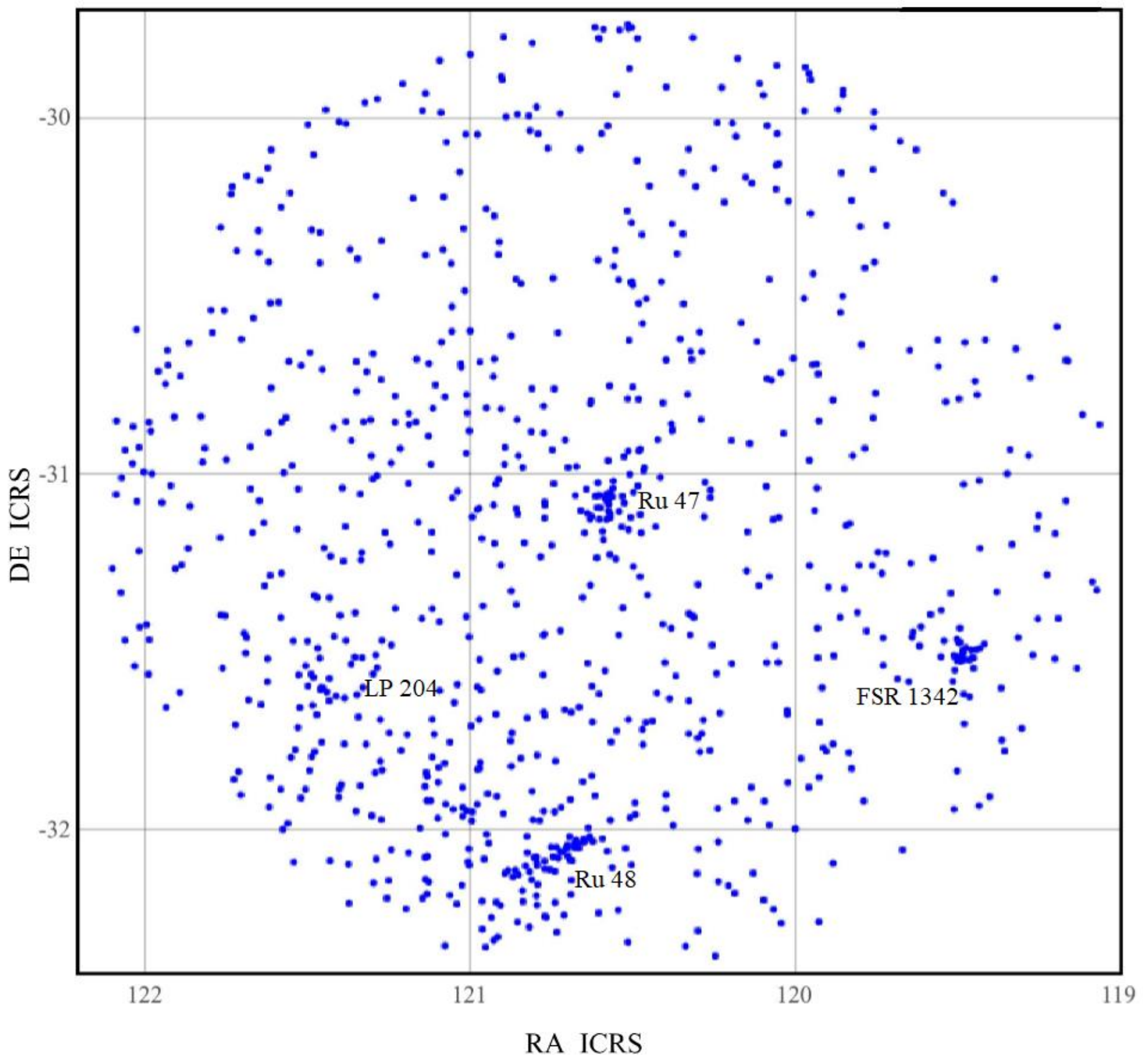

Figure 5. Chart of selected *Gaia* EDR3 sources around Ruprecht 47 (group 10). Constraints: *plx* 0.22 to 0.26 mas; $\mu_\alpha$ -2.3 to -2.6 mas $yr^{-1}$; $\mu_\delta$ 2.6 to 3.0 mas $yr^{-1}$.

The membership of Ruprecht 54 is also doubtful. The parallax and distance reported by Cantat-Gaudin et al. (2020) suggest that Ruprecht 54 might be far away than the rest of the siblings, even if the reevaluation of data from *Gaia* EDR3 reveals a parallax closer to them. Besides, the PMs are only marginally compatible, and it is standing apart from other group members. Moreover, it seems much older than most OCs of the system (Table 1), with age determinations ranging from 0.28 (Cantat-Gaudin et al. 2020) to 0.82 Gyr (Kharchenko et al. 2013). However, its only reported RV: 69 ± 16 km $s^{-1}$ (Tarricq et al. 2021) is compatible with RVs of Ruprecht 47: 60 km $s^{-1}$ (Dias et al. 2002) and 74 ± 2 km $s^{-1}$ (Tarricq et al. 2021) and the mentioned RV of LP 204. Unfortunately, there are no determinations of RV of the rest of the OCs to ascertain their membership in the group.

All data considered, the rest of the OCs of the group listed are firm candidates to be associated. Let's see the example of Ruprecht 47. Although its distance estimations (Table 1) seems somewhat small considering its parallax and the rest of the likely members, a distance of 4.4 ± 0.4 kpc has also been determined (Giorgi et al. 2015). While large deviations are found in the literature, most age estimations range from 0.03 Gyr (Table 1) to 0.08 Gyr (Dias et al. 2002; Giorgi et al. 2015), confirming that this is a young OC, as most of the probable members. Figure 5 shows a portrait of the most salient members of the group.

### Group 11

Kovaleva et al. (2020) have established that Collinder 135 and UBC 7 form a physical pair of OCs. Moreover, a string of stars seems to connect both clusters. According to Piskunov et al. (2006), this system is part of an extended open cluster complex associated with the Gould Belt, which includes 20 to 30 OCs as member candidates. Collinder 140, NGC 2451B and NGC 2547 have been proposed as members of the complex system (Soubiran et al. 2018; Beccari et al. 2020). Anyway, the present study has not found any other associated OCs in a field 600 arcmin wide around this binary system, probably due to their proximity to the sun (d ≤ 0.3 kpc).

The results reported in Table 1 confirm the case for a binary system. All reported parameters are within the adopted limits for a well-behaved pair: their real distance is much less than 0.1 kpc; the ratio Δplx/plx is less than 0.1 (the same is true when comparing photometric distances); the ratio ΔPM/plx is much lower than 2 (implying that the differences in tangential velocities are much lower than 10 km $s^{-1}$); the age ranges overlap, and the RVs match perfectly. Noticeably, the only significant differences are the computed positions of each OC centre from diverse authors.

### Group 12

Group 12 contains three member candidates (Figure 6), including two already known clusters: van den Bergh-Hagen 19 (BH 19) and UBC 471. BH 19 is relatively well studied. Retrieved distances range from 3.1 kpc (Dias et al. 2002) to 4.6 kpc (Cantat-Gaudin et al. 2020), but the reported mean parallax (0.19 mas, Table 1) suggested a larger distance. Revision of this OC using *Gaia* EDR3 led to a larger parallax (0.23 mas), which would correspond to a distance within the reported range (~ 4.3 kpc). A systematic shift on parallaxes has been observed in numerous instances in the present study. This shift could be due to a global offset of *Gaia* DR2 parallaxes, which are 0.029 mas too small on the whole (Lindegren et al. 2018). Reported ages for BH 19 range from 0.28 Gyr (Dias et al. 2002) to 0.57 Gyr (Kharchenko et al. 2013), i.e. it is probably not a young cluster. The only retrieved RV is 86 ± 2 km $s^{-1}$ (Soubiran et al. 2018).

UBC 471 has been less studied but presents the same mismatch between photometric distance (4.4 kpc; Cantat-Gaudin et al. 2020) and *Gaia* DR2 parallax (0.19 mas). This issue is again solved by the mean *Gaia* EDR3 parallax obtained in the present study, which suggests a distance of ~ 4.5 kpc. The only age reported for UBC 471 is 0.71 Gyr (Table 1) and we have obtained RV = 74 ± 3 km $s^{-1}$ (median of 3 member stars from *Gaia* data) in full agreement with the mean of 4 stars reported later in Tarricq et al. (2021). Despite a good match of position and kinematics (Table 1 and Fig. 6), some data of these OCs are only marginally compatible and further study of all candidate members of the group is necessary to confirm their membership. If true, this would be the first case of a triple system of mature OCs.

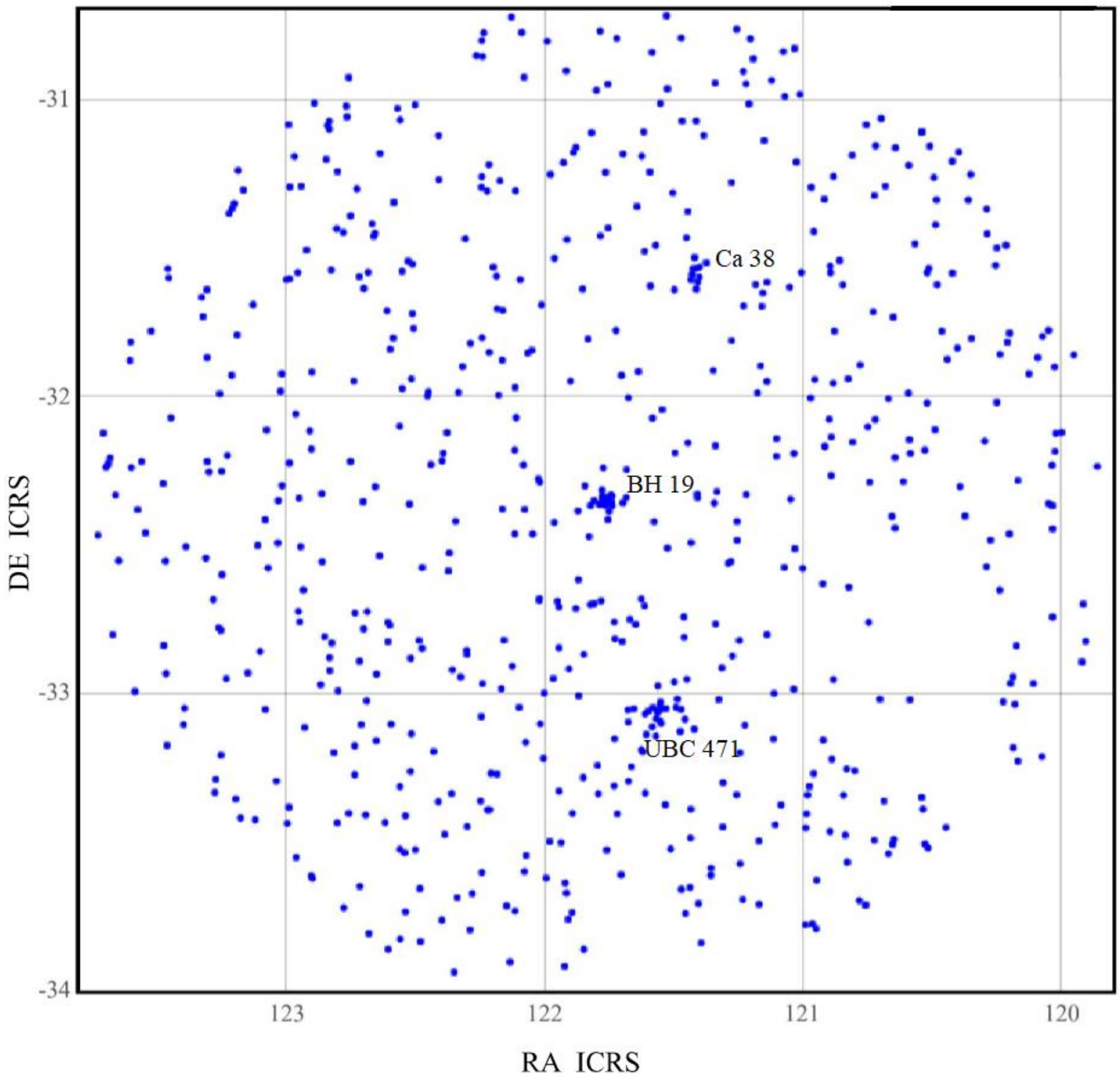

Figure 6. Chart of selected *Gaia* EDR3 sources around van den Bergh-Hagen 19 (Group 12). Constraints: *plx* 0.19 to 0.23 mas; $\mu_\alpha$ -2.7 to -3.0 mas yr$^{-1}$; $\mu_\delta$ 3.5 to 3.9 mas yr$^{-1}$; *Gmag* < 19.

## Group 13

This is a pair of two recently identified OCs: Casado 9 and 10 (Casado, 2020). Both are close in position and very similar in PMs and parallaxes. However, they are not a single spread cluster but a conspicuous double cluster. Two OC cores with a common corona are discernible in Figure 7.

To test the hypothesis of the double cluster, the CMD of the whole system (Figure 8) was built with all the sources of $G$ magnitude $\leq$ 18 that fulfil the constraints of both OC. Although some spreading of stars is noticed, both main sequences match perfectly, and the distances reported for individual stars are consistent with the mean parallaxes and distances in Table 1. However, the large uncertainty in turn-off point and stellar masses precluded the determination of a reliable age for this system. No RV for any of the member stars has been retrieved. Even so, the ensemble of results suggests that all stars of this double cluster probably have a common origin. Further details about this pair are available in Casado (2020).

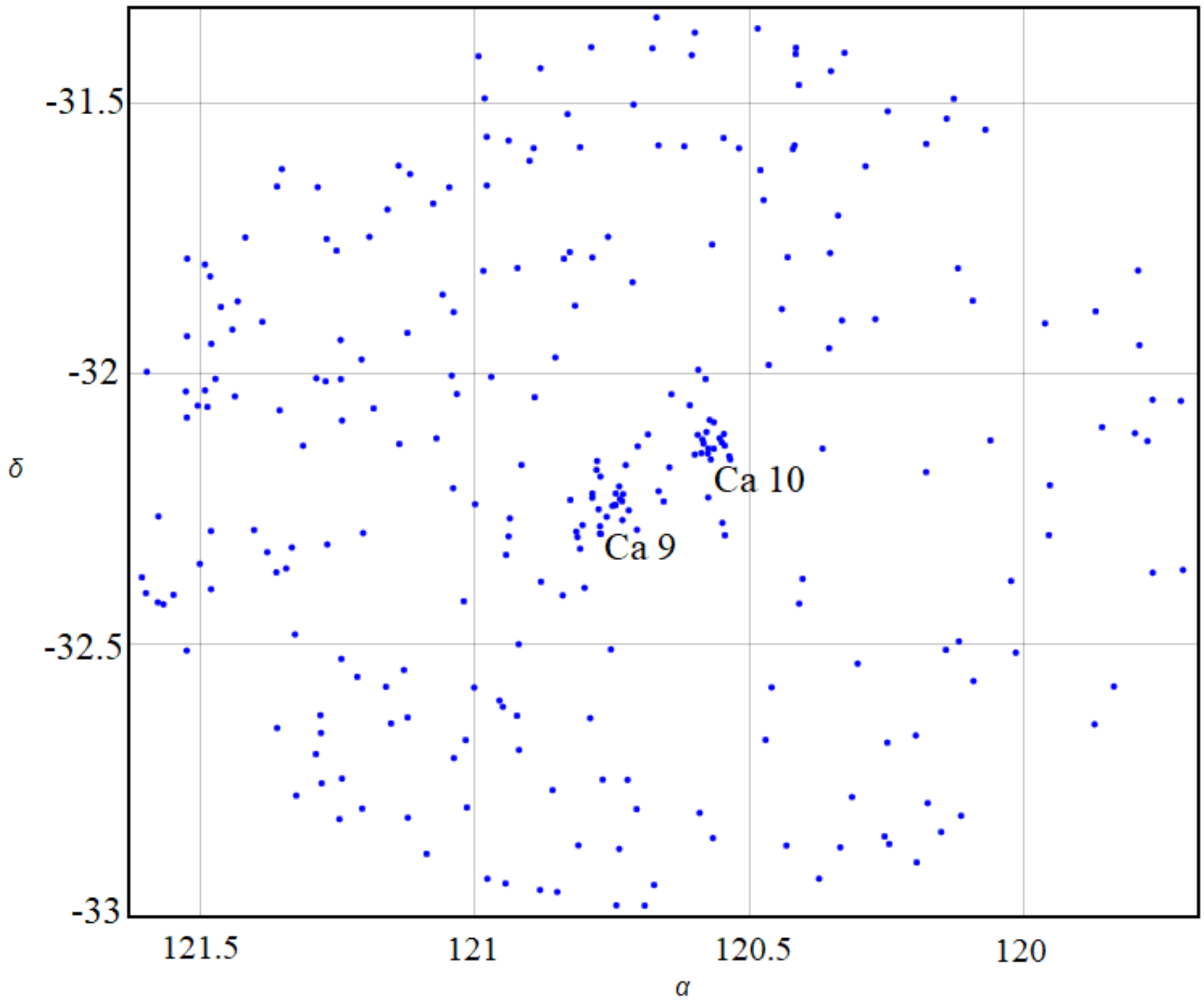

Figure 7. Location of likely star members of OCs Casado 9 and 10. Most noise from non-member field sources is removed by applying the same constraints as in Fig. 8.

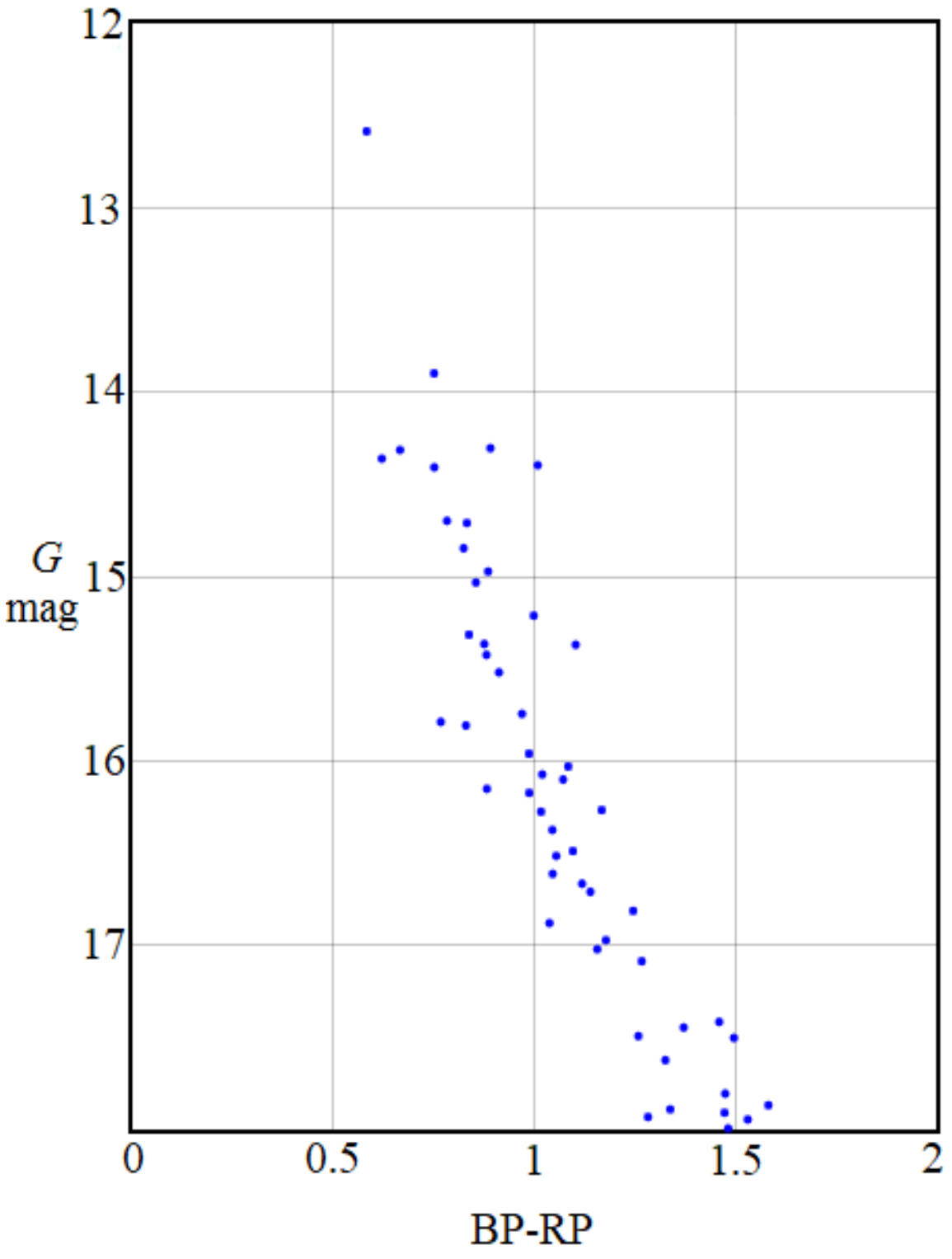

Figure 8. Distribution of the probable members of the double cluster comprising Casado 9 and 10 in the *G* magnitude versus BP-RP colour plane. Constraints: *plx* 0.22 to 0.28 mas; $\mu_\alpha$ -2.9 to -3.5 mas yr$^{-1}$; $\mu_\delta$ 3.5 to 3.9 mas yr$^{-1}$.

## *Group 14*

The close pair formed by UBC 644 and UBC 473 also deserves discussion. Once more, the positions, distances and kinematics are compatible within the reported error margins, but the ages are not (Table 1). However, different ages are not enough to rule out that two OCs form a pair, especially in this case, considering the small number of stars forming UBC 644 (the surface density of member stars is only slightly larger than the surrounding field). For UBC 473 we retrieved RV = 77 km $s^{-1}$ based on two stars, a result confirmed in Tarricq et al. (2021), but no RV has been found for UBC 644. Our approach has been again to obtain a combined CMD for the system of both OCs. In this case, the resulting CMD showed a too broad main sequence to infer a common origin for both clusters, in agreement with the different reported ages. Thus, this pair is considered doubtful. Nevertheless, a gravitational capture cannot be ruled out at present. So, further work is needed to conclusively confirm or discard the existence of this double cluster. No other OCs sharing compatible PMs and parallaxes have been found in a field 300 arcmin wide around this candidate system.

## *Group 15*

Group 15 candidates include two known OCs and a new one. All of them seem to be young since they still appear associated with the cloud where they were presumably formed.

*Gaia* data on MWSC 1460 suggest its likely membership to the group. It has a double core, its reported distance is 3.0 kpc (compatible with *Gaia* EDR3 parallax) and its age is ca. 0.013 Gyr (Dias et al. 2002). RV is 43 ± 6 km $s^{-1}$ (Conrad et al. 2017). Casado 24 was found during the study of this group and seems to be associated with the adjacent HII region BRAN 133, which suggests it is also young. Its derived parameters look like a twin of MWSC 1460, which endorses its likely membership.

On the other hand, the adscription of Majaess 95 to the group is uncertain despite its central position. This is one of those mentioned embedded clusters or protoclusters with a few stars well measured by *Gaia* (10 sources of $G < 18$). We have tried our best to estimate its mean characteristics, but the error bars are higher than usual and the results, especially the parallax, are only marginally compatible with other members of the group. The resulting CMD is ill-defined. Perhaps these are the reasons why no value of its distance, parallax, age, and RV based on its member stars had been reported so far. However, this cluster has been associated with the star-forming region 254.05-0.10 (Avedisova 2002), which has a measured RV of 31 km $s^{-1}$. Unfortunately, this RV is not decisive on the membership of Majaess 95 to the group. No other possibly associated OC was found in a circular area of radius 100 arcmin centered in the position of Majaess 95.

## *Group 16*

The binary system formed by Collinder 197 and MWSC 1579 was formerly proposed by de La Fuente Marcos & de La Fuente Marcos (2009). Reported RVs of Collinder 197 are 31± 9 km $s^{-1}$ (Dias et al. 2002), 33 ± 4 km $s^{-1}$ (Kharchenko et al. 2013; Loktin & Popova 2017) and 36 ± 3 km $s^{-1}$ (Soubiran et al. 2018). However, the ultimate RV seems to be 20.8 ± 0.5 km $s^{-1}$ (Tarricq et al. 2021) obtained from the mean of 101 stars. For MWSC 1579, the only reported RV is 17 ± 9 km $s^{-1}$ (Kharchenko et al. 2013; Loktin & Popova 2017). The RVs of both OCs are thus compatible, reinforcing the case for a binary system since the rest of their parameters are consistent (with minor dispersion in ages from different authors, as usual).

Two other member candidates have been identified in the surrounding field, namely Ruprecht 64 and Pismis 5. We have reexamined Ruprecht 64 using *Gaia* EDR3. Its updated parameters (Table 1) seem to nicely fit those of the previously identified pair, despite its reported ages span from 0.07 Gyr (Loktin & Popova 2017) to 0.28 Gyr (Kharchenko et al. 2013). Reported RVs, from 21 km $s^{-1}$ (Kharchenko et al. 2005) to 27 ± 6 km $s^{-1}$ (Loktin & Popova 2017), are also compatible with the rest of the group. The membership of Pismis 5 is less clear, however. Although most of its reported parameters, including its age, are well-matched, $\mu_\delta$ seems only marginally compatible, and there is no reported RV to verify its affiliation to the group.

### *Group 17*

Ruprecht 65 and UBC 245 appear to form a textbook example of a primordial pair. Not only their *Gaia* data fit almost perfectly, but their CMDs and the resulting ages present a good match too.

### *Group 18*

Liu and Pang (2019) identified a group of 4 OCs, namely NGC 2659, Pismis 8, Ruprecht 71, and LP 58. However, the parameters available for Ruprecht 71 could best fit in group 22. Conversely, we have found some extra candidate members for this group, including NGC 2645, Gulliver 5 and two previously unknown OCs (Table 1). The pair formed by NGC 2645 and Pismis 8 was first proposed by Subramanian et al. (1995).

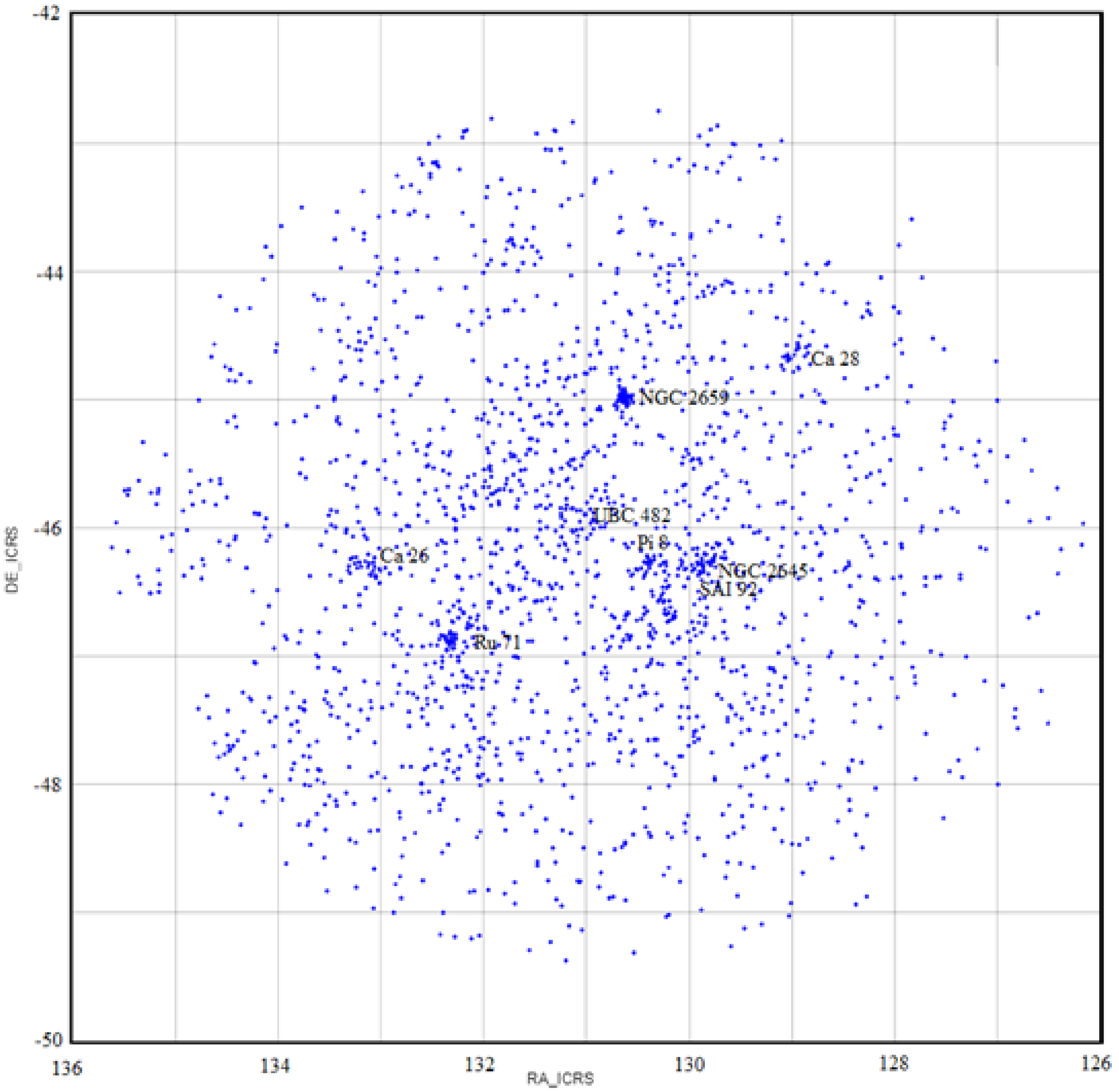

Figure 9. Chart of selected *Gaia* EDR3 sources of group 18. Constraints: *plx* 0.44 to 0.53 mas; $\mu_\alpha$ -5.1 to -6.0 mas yr$^{-1}$; $\mu_\delta$ 4.5 to 5.2 mas yr$^{-1}$; *Gmag* < 17. A common halo of stars can be noticed between most of the OCs in the central area of the image. Two OCs of group 22 are also spotted.

The reexamination of Gulliver 5 led to significantly larger parallax and counts of member stars, but the astrometric data are only marginally compatible with some of the group member candidates. Similarly, the ensemble of data of NGC 2659, UBC 482 and Casado 61 does not allow claiming a clear membership of these OCs. For instance, all the mean parallaxes of these four OCs lie in the range of *plx* = 0.41-0.45 mas, while those of the rest of the candidate members lie in the range of *plx* = 0.51-0.54 mas. Similar grouping is found in $\mu_\alpha$, with mean values from -5.0 to -5.3 mas yr$^{-1}$ for this possible subgroup and from -5.5 to -5.9 mas yr$^{-1}$ for the rest of the candidate members. In any case, the available data on these four OCs suggest that they form a probable physical system, regardless it belongs or not to the whole group.

The association of NGC 2645 and SAI 92 (or FSR 1436), proposed by Bica et al. (2019), was also dubious because few parameters of the second had been reported and ages are not precise enough, ranging from ~ 0.10 to ~ 0.01 Gyr (Glushkova et al. 2010; Buckner & Froebrich 2014). However, the results obtained in the present study allow establishing its likely membership to the same group (Table 1 and Figure 9). The proximity of SAI 92 and LP 58 compared to their radii (less than 5 arcmin apart, unresolved in Fig. 9), and the good match of the rest of their characteristics suggest they are probably the same OC.

Reported RVs are scarce but enlightening. RVs match for NGC 2645 and NGC 2659: 17 ± 4 km s$^{-1}$ (Kharchenko et al. 2013; Loktin & Popova 2017) and 17 ± 2 km s$^{-1}$ (Soubiran et al. 2018; Tarricq et al. 2021), respectively, which strengths the likelihood of their link (despite their *plx* and $\mu_\alpha$ are only marginally compatible). However, the only determination of RV from a single star of Pismis 8 yields a value of 63 ± 10 km s$^{-1}$ (Dias et al. 2002), which casts some doubt on the affiliation of this OC to the same group.

### *Group 19*

The reported average positions of both OCs in this group are 16 arcmin apart, which is much smaller than the maximum radius of LP 2388 (71 arcmin), i. e. both clusters mostly overlap. The rest of the astrometric parameters match almost perfectly. Only ages differ significantly (Table 1), but the number of adopted star members of UBC 480 (13) seems too few to obtain an accurate age. The different number of stars in each case could be related to the diverse allowed uncertainties in PM and parallax. Accordingly, it is observed that the OCs described in Liu & Pang (2019) are generally larger and more populated than the same objects in Cantat-Gaudin et al. (2020). All things considered, our primary hypothesis is that LP 2388 and UBC 480 are the same OC. For this system, we have found a probable member (*Gaia* source 5513999582929392384) having RV = 19 ± 9 km s$^{-1}$. No associated OCs have been found in a circle of radius 400 arcmin around LP 2388.

### *Group 20*

Bochum 7 appears to be near Casado 46, and the structure of this subsystem, including multiple cores, suggest a probable link between both members. All reported distances to Bochum 7 agree: 5.9 ± 0.2 kpc (Dias et al. 2002; Kharchenko et al. 2013) and are consistent with the measured parallaxes of the rest of the group members (despite parallaxes are not precise for $d > 5$ kpc). Bochum 7 is young, with ages that range from 0.01 Gyr (Dias et al. 2002) to 0.03 Gyr (Kharchenko et al. 2013). Casado 46 could be young also, given its associated nebulosity and its similar CMD. Considering only their overlapping if the classical large radius of Bochum 7 (33.5 pc; van den Bergh 2006) is assumed, it is tempting to conclude that the new OC is part of Bochum 7. Nevertheless, both objects are well detached (Figure 10), and their CMDs do not perfectly fit.

On the other hand, the CMDs of Casado 48 and Teutsch 65 are well-matched, suggesting some affiliation among them. However, the distances between Teutsch 65 and Casado 48 (as well as for Teutsch 65 and Bochum 7) are slightly farther than 0.1 kpc, making dubious their link. The membership of Casado 48 to the group is even less likely because of some deviation in $\mu_\alpha$. Notice that a difference of 0.3 mas yr$^{-1}$ at such a heliocentric distance implies a difference in PM rate of more than 10 km $s^{-1}$. Unfortunately, no RVs for Teutsch 65 or Casado 48 have been retrieved to figure out these doubts.

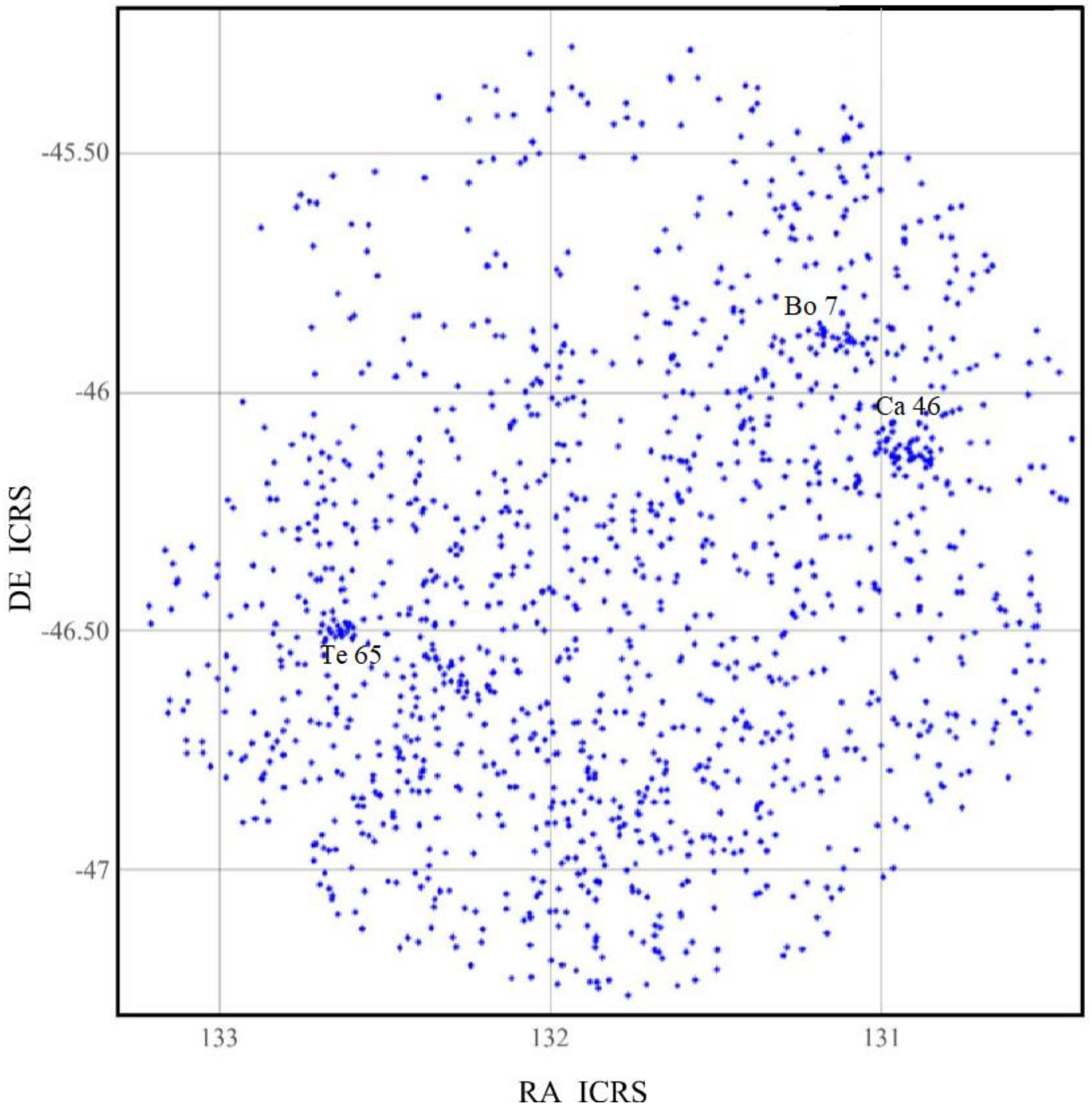

Figure 10. Chart of selected *Gaia* EDR3 sources of group 20. Constraints: *plx* 0.12 to 0.22 mas; $\mu_\alpha$ -2.9 to -3.3 mas $yr^{-1}$; $\mu_\delta$ 3.4 to 3.8 mas $yr^{-1}$; *Gmag* < 18.

The membership of Casado 49 is dubious due also to a similar deviation of $\mu_\alpha$, despite all the new OCs were found during the study of the group because they share comparable astrometric data. On the other hand, the RV of one of its core stars agrees with the reported RV of Bochum 7 (Table 1). Anyway, some link between Casado 49 and Teutsch 222 seems also likely, even in the case they form a pair unrelated to the rest of the group.

### *Group 21*

The membership of Casado 53 and 58 to the same group needs confirmation. Although their positions (13 arcmin apart) and astrometric data are close enough, their raw CMDs are not well-matched. Such a deficient fit could be due, however, to their different extinction and reddening in a field with variable nebulosity. Interstellar clouds are more evident around Casado 58, which is likely to be a young OC.

Casado 59, at 0.08 kpc of Casado 53 (assuming that both are at the same heliocentric distance), holds compatible measurements with its siblings. However, the membership to the same group would require confirmation through further data on all candidates. *Gaia* DR2 RV of one of the sources of Casado 59 suggests a preliminary value of 70 ± 1 km s$^{-1}$ for this OC. A new OC remnant candidate with a few clumped stars at galactic coordinates 264.705, -2.003 could be also part of this system.

This group has similar astrometric parameters to group 20 and both systems might be related.

### Group 22

Group 22 might be a subsystem of group 18 (see Fig. 9). All characteristics agree, except $\mu_\delta$, which are somewhat smaller. The candidate members of this group are depicted in Figure 11.

There is a wide range of reported ages for Ruprecht 71. Moreover, the same catalogue proposes two incongruent ages: ca. 0.02 and 1.05 Gyr (Liu & Pang 2019). These divergences suggest the possibility that we are observing two OCs in the line-of-sight of Ruprecht 71 (possibly along with Ruprecht 70). Anyway, other authors converge in 0.02 Gyr (Cantat-Gaudin et al. 2020) for the specific OC defined by the parameters in Table 1. The only reported RV is 39 ± 1 km s$^{-1}$ (Soubiran et al. 2018), from just one *Gaia* source. A bridge of halo stars, including a small group of six sources at galactic coordinates 267.14, -1.76, seems to connect this OC to the candidate companion UBC 483 (Fig. 11). Both likely members share well-matched characteristics, although Casado 26 appears closer to Ruprecht 71 than UBC 483.

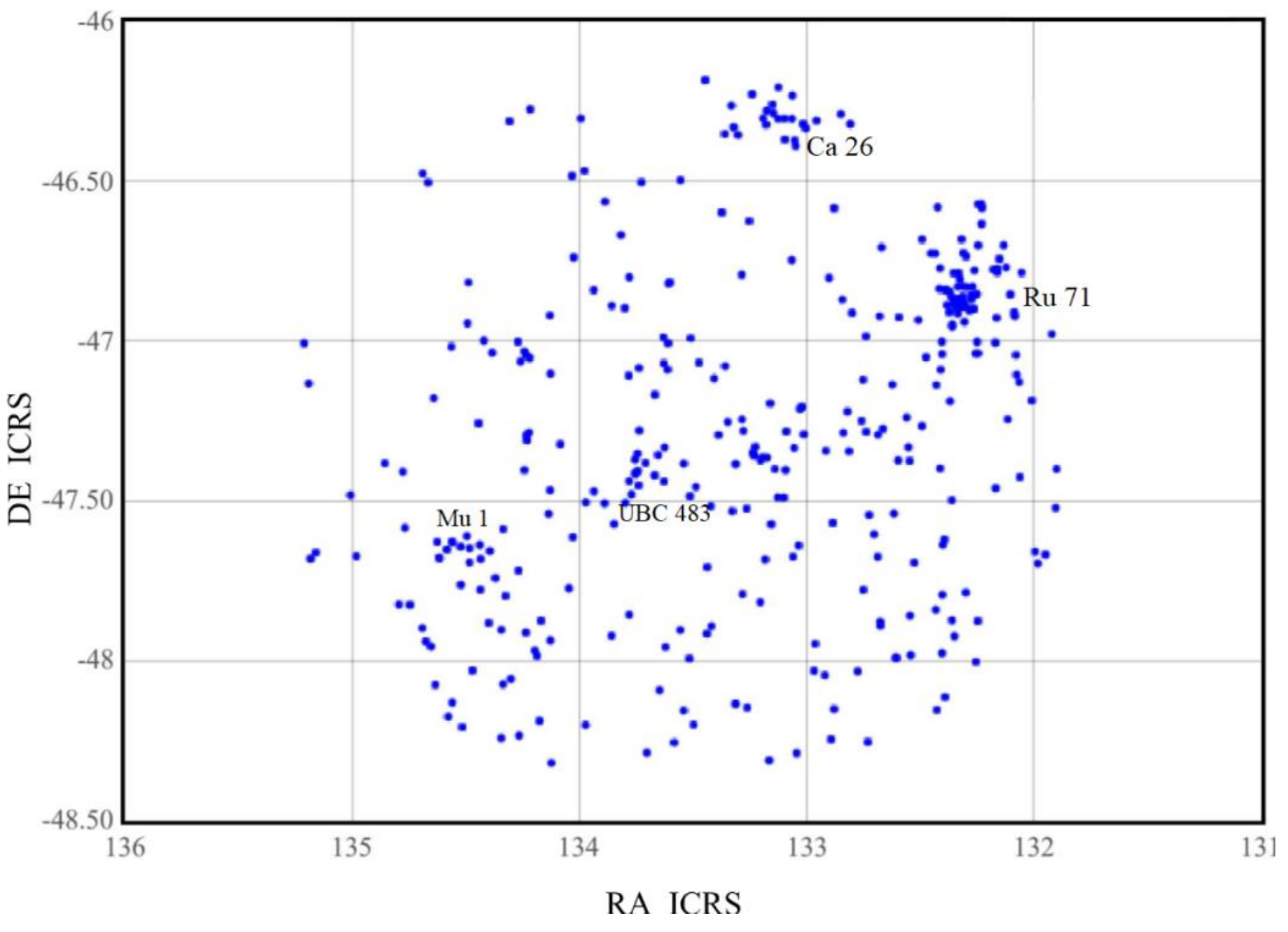

Figure 11. Chart of selected *Gaia* EDR3 sources defining the group of Ruprecht 71 (group 22). Constraints: *plx* 0.45 to 0.53 mas; $\mu_\alpha$ -5.0 to -5.4 mas yr$^{-1}$; $\mu_\delta$ 4.4 to 4.8 mas yr$^{-1}$; *Gmag* < 19.

The membership of Muzzio 1 is less clear, however. All its reported ages range from 0.003 Gyr (Kharchenko et al. 2013) to 0.008 Gyr (Cantat-Gaudin et al. 2020), i.e. Muzzio 1 is a very young cluster as well. But a reported RV of 136 ± 6 km s$^{-1}$ (Soubiran et al. 2018) would rule out the alleged relationship with the group, if confirmed. However, we should be cautious because there is a very different RV: -13 km s$^{-1}$ (Kharchenko et al. 2013). Moreover, each RV comes from only one star. Unfortunately, we have not found in our revision any member star having RV in *Gaia* DR2.

### Group 23

All *Gaia* data on this pair of previously undetected OCs are consistent. Two stars with marginally compatible RVs were found (one per cluster; Table 1). No other linked OCs were found in fields of radii 100 arcmin around this likely double cluster.

### Group 24

ESO 312-4 seems to form a pair with the new OC Casado 60 (Table 1). Both are almost 20 arcmin apart, which corresponds to ca. 40 pc at their heliocentric distance. The mean parallaxes of both clusters seem only marginally compatible at first sight. However, considering the offset of *Gaia* DR2 parallaxes, a corrected value of 0.18 mas for ESO 312-4 seems more accurate since it is compatible with its photometric distance (Table 1) and with the *Gaia* EDR3 mean parallax of Casado 60. The rest of the parameters of this likely pair are well-matched.

There is, once more, a wide span in the reported ages of ESO 312-4: from 0.04 Gyr (Cantat-Gaudin et al. 2020) to 0.56 Gyr (Dias et al. 2002). However, the most recent determination suggests that it is a young OC, as usually found in this study.

A small clump (0.7 arcmin wide) of 8 stars at galactic coordinates 259.331, -1.793, which shares the same astrometric parameters, also seems to belong to this group.

## 5. A statistical approach

An overview of the ensemble of data compiled in Table 1 allows counting a total of 22 candidate groups (two of them have been ruled out) comprising 80 possible members. These counts lead to an average population of 3.6 members per group. Considering only the most likely members and their groups, the average population decreases to 52/17 = 3.1 OCs per group. Notice also that seven of these OCs have double cores. Eight out of these 17 groups are double clusters (considering only the probable members). The most populated groups (#9 and #18) count up to eight and nine OCs respectively, and seven of them are likely members. The relatively high population of OCs per group suggests a common origin for most of the groups since a gravitational capture of more than two OCs seems exceedingly improbable.

As mentioned above, all except one of the likely candidate members are young, and no convincing case for a physical pair/group of OCs of ages > 0.1 Gyr has been found. These results also point to a low probability, if any, of pairs formed by tidal capture or resonant trapping, which would be due to the small likelihood of close encounters of OCs, and the even lower probability of tidal capture without disruption of at least one of the clusters. Our statistics contrast with de La Fuente Marcos & de La Fuente Marcos (2009), who stated that only ~ 40% of pairs are probably primordial. Therefore, we have revisited their work, and we found again that all their candidate pairs of OCs older than 0.1 Gyr are not physically linked (to be published in a forthcoming paper). Congruently, the age of most of the discarded candidates in the present study is more than 0.1 Gyr. Our findings agree, at least qualitatively, with those of the pioneering work of Larsen (2004), who studied young OCs in nearby spiral galaxies, and found that many of the youngest objects are in very crowded regions, and about 1/3–1/2 of them are double or multiple sources.

Of the 52 most likely members, only 14 have $b > 0$. 21 of them are in the range of parallax from 0.20 to 0.30 mas, which merely reflects the peak frequency of *Gaia* sources in the cited interval. On the other hand, only five of them have $plx > 1$ mas. As expected, the uncertainties in distances obtained from parallax grow faster than those obtained photometrically and become predominant at $plx \leq 0.2$. All $\mu_\alpha$ are negative and all $\mu_\delta$ are positive, reflecting the mainstream motion of the stars in the studied part of the galactic disk. In line with that, all of the candidates having RVs seem to be receding from the sun.

We have used the following procedure to obtain a statistical estimate on the fraction of OCs that form double and multiple systems: First, we take as reference the most recent of the catalogues used in these work, namely that of Cantat-Gaudin et al. (2020), mainly because it is the only comprehensive enough catalogue (2017 OCs, 206 of which in the 30º sector of the Galaxy studied in here) that contains the set of recently discovered UBC clusters. Note that a significant number of the listed OCs in Table 1 are UBC clusters. Then, we count the total number of member candidates listed in the catalogue (see column C20 in Table 1), and we have 42 OCs, i.e. ~20 % of the sample. The fraction decreases to ~15 % of the compiled OCs if we consider only the most likely members of the groups (31 OCs in the cited catalogue).

Of course, this is not a definitive determination of the fraction of binary/multiple OCs in the Galactic disk because of the incompleteness of the analyzed samples. However, the obtained fractions fit comfortably within the 8% and 20% margins mentioned in the introduction. Moreover, considering that the studied sector contains the Vela-Puppis star formation region with a young OCs population above average, the obtained 15% fraction is compatible with the hypothesis that the actual fraction is similar to that in the LMC (de La Fuente Marcos & de La Fuente Marcos 2009). Similar statistics considering only the 202 OCs listed in Bica et al. (2019) for the studied sector of the Galaxy (column B19 in Table 1), yields somewhat smaller figures: ~ 13% and ~ 9.4% for all the candidates and the probable candidates, respectively. Anyway, these values are also compatible with the same hypothesis stressing the resemblance of the Galaxy and the LMC in this respect.

## 6. Concluding remarks

The vast majority of the detected groups are formed by clusters younger than 0.1 Gyr, and no plausible case for a group formed by older OCs has been found. The average population is three to four members per group. These results indicate that most groups are of primordial origin and are not stable for a long time, in line with similar conclusions obtained from the study of the Magellanic Clouds (Hatzidimitriou & Bhatia 1990; Dieball et al. 2002). OCs with double cores, possibly experiencing fragmentation, are not rare in the studied sample. For all these reasons, the present findings support that OCs are born in groups like stars are born in clusters (see also de La Fuente Marcos & de La Fuente Marcos 2009b).

The only two pairs discarded in Table 1 –namely, those formed by UBC 637/UBC 638 and LP 2388/UBC 480- have been ruled out because the members of each pair are probably part of the same OC. SAI 92 and LP 58 may also be a single OC, and Waterloo 3 seems to be part of UBC 465.

The census of double/multiple candidate OCs here described has increased considerably versus previous reports. The relatively large number of likely candidates casts doubt on the statement that double clusters form preferentially in environments different to that of the third Galactic quadrant (Vázquez et al. 2010). A preliminary estimation of the fraction of potential members of the groups here reported versus the general OC population (ca. 9.4 to 15%) is similar to the proportion observed in the LMC.

The hunting for associated clusters around a given group is an excellent method of discovering new OCs, in part because the density of OCs around the groups is naturally higher than average. Moreover, the PMs and parallaxes of the former group members are a guide for a search that, otherwise, is practically blind, i.e. we cannot see the wood for the trees. Fourteen out of 24 new OCs found in this way are likely members of the studied groups.

On the other hand, accurate age would help to confirm the possible membership of a part of the group candidate members. More RV measurements are also required to clarify the affiliation of the dubious members to the groups and compute the orbits of OCs around the galactic centre. Tarricq et al. (2021) have given a significant step forward in this way very recently, mainly based on the mining of the *Gaia* data.

## Acknowledgments

This work has made use of data from the European Space Agency (ESA) mission *Gaia* (https://www.cosmos.esa.int/gaia), processed by the *Gaia* Data Processing and Analysis Consortium (DPAC, https://www.cosmos.esa.int/web/gaia/dpac/consortium). Funding for the DPAC has been provided by national institutions, in particular the institutions participating in the *Gaia* Multilateral Agreement. This research made extensive use of the SIMBAD database, and the VizieR catalogue access tool, operated at the CDS, Strasbourg, France (DOI: 10.26093/cds/vizier), and of NASA Astrophysics Data System Bibliographic Services.